\documentclass[twocolumn]{aastex631}

\usepackage{microtype}  
\usepackage{amsmath}
\usepackage{amsfonts}
\usepackage{amssymb}
\usepackage{multirow}

\newcommand{\mlg}{\ensuremath{M_{\rm LG}}}
\newcommand{\mmto}{\ensuremath{M_{\rm M31}}}
\newcommand{\mmw}{\ensuremath{M_{\rm MW}}}
\newcommand{\vtan}{\ensuremath{v_\textrm{tan}}}
\newcommand{\vrad}{\ensuremath{v_\textrm{rad}}}

\newcommand{\bov}{\ensuremath{\boldsymbol{v}}}
\newcommand{\boldx}{\ensuremath{\boldsymbol{x}}}
\newcommand{\vtrav}{\ensuremath{\bov_{\rm travel}}}
\newcommand{\xtrav}{\ensuremath{\boldx_{\rm travel}}}
\newcommand{\pos}[2]{\ensuremath{\boldx_{\rm #1 \to #2}}}
\newcommand{\vel}[2]{\ensuremath{\bov_{\rm #1 \to #2}}}

\newcommand{\mwouter}{\ensuremath{\textrm{MW}_\textrm{halo}}}
\newcommand{\mwdisk}{\ensuremath{\textrm{MW}_\textrm{disk}}}
\newcommand{\reflabel}[1]{\ensuremath{^{\mbox{\scriptsize{#1}}}}}

\shorttitle{Updated Local Group Mass from Timing Argument}
\shortauthors{Chamberlain et al.}

\graphicspath{{./}}

\renewcommand{\vec}[1]{\ensuremath{\bs{#1}}}

\newcommand{\Msun}{\ensuremath{\mathrm{M}_\odot}}

\newcommand{\kms}{\ensuremath{\mathrm{km}~\mathrm{s}^{-1}}}

\newcommand{\kpc}{\ensuremath{\mathrm{kpc}}}

\newcommand{\Gyr}{\ensuremath{\mathrm{Gyr}}}

\newcommand{\masyr}{\ensuremath{\mathrm{mas}~\mathrm{yr}^{-1}}}
\newcommand{\muasyr}{\ensuremath{\mu\mathrm{as}~\mathrm{yr}^{-1}}}

\newcommand{\bs}[1]{\boldsymbol{#1}}

\newcommand{\gaia}{\textsl{Gaia}}

\newcommand{\hst}{\textsl{HST}}

\newcommand{\affuofa}{University of Arizona, 933 N. Cherry Ave,
    Tucson, AZ 85721, USA}
\newcommand{\affcca}{Center for Computational Astrophysics, Flatiron Institute,
    Simons Foundation, 162 Fifth Avenue, New York, NY 10010, USA}
\newcommand{\affedinb}{Institute for Astronomy, University of Edinburgh,
    Royal Observatory, Blackford Hill, Edinburgh EH9 3HJ, UK}
\newcommand{\affparis}{Institut d'Astrophysique de Paris,
    98 Bis Blvd Arago 75014 Paris, France}

\begin{document}

\title{
  Implications of the Milky Way travel velocity for dynamical\\
  mass estimates of the Local Group
}

\author[0000-0001-8765-8670]{Katie~Chamberlain}
\affiliation{\affuofa}
\affiliation{\affcca}

\author[0000-0003-0872-7098]{Adrian~M.~Price-Whelan}
\affiliation{\affcca}

\author[0000-0003-0715-2173]{Gurtina Besla}
\affiliation{\affuofa}

\author[0000-0002-6993-0826]{Emily C. Cunningham}
\affiliation{\affcca}

\author[0000-0001-7107-1744]{Nicol\'{a}s Garavito-Camargo}
\affiliation{\affcca}

\author{Jorge Pe\~{n}arrubia}
\affiliation{\affedinb}

\author[0000-0003-1517-3935 ]{Michael S. Petersen}
\affiliation{\affparis}

\begin{abstract}
  The total mass of the Local Group (LG) is a fundamental quantity that enables interpreting the orbits of its constituent galaxies and placing the LG in a cosmological context. 
  One of the few methods that allows inferring the total mass directly is the ``Timing Argument,'' which models the relative orbit of the Milky Way (MW) and M31 in equilibrium. 
  The MW itself is not in equilibrium, a byproduct of its merger history including the recent pericentric passage of the LMC, and recent work has found that the MW disk is moving with a lower bound ``travel velocity'' of $\sim 32~\kms$ with respect to the outer stellar halo.
  Previous Timing Argument measurements attempt to account for this non-equilibrium state, but have been restricted to theoretical predictions for the impact of the LMC specifically.
  In this paper, we quantify the impact of a travel velocity on recovered LG mass estimates using several different compilations of recent kinematic measurements of M31. 
  We find that incorporating the measured value of the travel velocity lowers the inferred LG mass by 10--12\% compared to a static MW halo.
  Measurements of the travel velocity with more distant tracers could yield even larger values, which would further decrease the inferred LG mass. 
  Therefore, the newly measured travel velocity directly implies a lower LG mass than from a model with a static MW halo and must be considered in future dynamical studies of the Local Volume.
\end{abstract}

\section{Introduction}
\label{sec:intro}
The total mass of the Local Group (LG) is an important quantity in many
local cosmological and Milky Way (MW) applications.
For example, it is used to identify analogous halos in cosmological simulations
and thus allows comparing host galaxy and satellite galaxy number counts and
properties \citep[e.g.,][]{Marinacci:2017, Dooley2017, Patel2017a, Besla2018,
Patel2018, Garrison-Kimmel:2019a, Garrison-Kimmel:2019b, Sawala2022}.
It is also used to turn the kinematics of LG galaxies into orbital histories \citep[e.g.,][]{Peebles:2017}, which is used to
interpret their gas content \citep[e.g.,][]{Fillingham:2018,Putman:2021} and
star formation histories \citep[e.g.,][]{Tolstoy:2009}.
However, as most of the mass in the LG is in dark matter distributed over megaparsec scales, it is difficult to directly measure its total mass.

Given its utility in studies of the local universe, several methods have been
used to dynamically infer the mass of the LG.
Many of these techniques determine the individual masses of the MW and M31
independently \citep[e.g.][]{Watkins2010, Fardal2013,Diaz2014,Carlesi2017,
Patel2018, Eadie:2019, Fritz:2020, Deason:2021, Villanueva-Domingo2021,
Wang:2022}, often via the dynamics of their satellites
and stellar streams, then combine them to get an estimate of the total LG mass.
However, these methods generally only measure the enclosed mass of the MW or M31
within some internal radius (i.e., much smaller than LG scales) and then
extrapolate, leading to mass-profile-dependent estimates of the total LG mass.
Other techniques aim to more directly measure the mass of the LG en masse, for
example looking for Local Group analogs in cosmological simulations based on stellar mass and kinematic criteria
\citep[e.g.,][]{LiWhite2008, Gonzalez2014, Zhai2020, Hartl2022},
by studying the kinematics of Local Volume (LV) galaxies
\citep[e.g.,][]{Diaz2014,Penarrubia2014}, or by applying machine learning (ML)
techniques to hydrodynamic simulation data
\citep[e.g.,][]{McLeod2017,Villanueva-Domingo2021}.
One of the earliest methods utilized in this vein is the ``Timing Argument,''
which uses the fact that the LG galaxies (most often the MW and M31) are bound and approaching pericenter in their relative orbit, but must have been close enough over cosmic time to not be pulled apart by the Hubble flow.
The Timing Argument can be generalized to simultaneously model the orbits of LV galaxies around the LG
~\citep{Penarrubia2016, Penarrubia2017}, but here we restrict our analysis to
the ``classic'' Timing Argument using only the MW and M31.
We summarize the relevant details of the Timing Argument method in
Section~\ref{sec:timingarg}.

The Timing Argument (using the MW and M31) uses the observed kinematics of M31
to model the relative orbit of the two galaxies as a Keplerian orbit.
Assuming Keplerian dynamics enables dynamically measuring the total mass of the
MW and M31 with analytic expressions for all relevant kinematic quantities
because of the simplicity of the two-body equations of motion.
The inferred mass from the Timing Argument thus directly depends on the observationally-measured kinematics of the M31 center.

However, the LG is not in equilibrium.
In the past decade, a number of studies have begun to consider of the
impact of the Large Magellanic Cloud (LMC) on the mass and inferred
dynamics of the LG.
\citet{Penarrubia2016} studied the effect of the presence of the LMC on the total
mass estimates of the LG via the Timing Argument by modelling the motion of M31 about the MW-LMC barycenter, and using the
kinematics of 35 LV Galaxies to simultaneously measure
a MW mass of $\rm M_{MW} = 1.04^{+0.26}_{-0.23}\times 10^{12} M_\odot$,
M31 mass of $\rm M_{M31} = 1.33^{+0.39}_{-0.33}\times 10^{12} M_\odot$,
LMC mass of $\rm M_{LMC} = 0.25^{+0.09}_{-0.08}\times 10^{12} M_\odot$,
and LG mass of $\rm M_{LG} = 2.64^{+0.42}_{-0.38}\times 10^{12}
M_\odot$.
Another recent Timing Argument work by \cite{Benisty2022} modelled the orbital
history of M31 and the MW, with and without a mass and orbital model of the LMC,
to estimate the contribution of the LMC-induced shift in the MW barycenter on
the measured tangential and radial velocities of M31, then applied these
corrections to their model to remove the impact of the LMC in their analysis,
and found that the inferred LG mass decreased by ~10\%.
 
Recent studies of the dynamics of the MW and its satellites have
revealed that the infall of the Magellanic Clouds (MCs) is
causing significant distortions to the dark matter and stellar distribution in
the MW halo \citep{Laporte:2018a, Laporte:2018b, Garavito-Camargo:2019,
Conroy:2021, Erkal:2021}.
In addition, numerous studies of the interaction between the MW and LMC (using
simulated analogs) have quantified the expected LMC-induced reflex motion of the
MW 
disk and inner halo, which are likely being accelerated away from the center-of-mass reference frame of a static MW halo \citep{Gomez2015, Cunningham:2020,
Petersen:2020, Garavito-Camargo2021b}.
The induced systematic shift in the measurements of the M31 kinematics may have created a bias in previous mass measurements via the Timing Argument, thus
impacting interpretations of LG dynamics, orbital histories, cosmological
context, etc.

Previous Timing Argument studies have accounted for the impact of an LMC-induced reflex motion on the orbital histories of the other galaxies in the LG.
However, other satellite mergers such as the ongoing merger with the
Sagittarius dwarf galaxy, as well as the past merger with the 
progenitor of the Gaia--Enceladus--Sausage, have likely also imparted their own reflex motion to the inner MW halo.
A signature of the reflex motion of the MW is imprinted as a velocity dipole in the radial velocities of stars in the outer stellar halo
\cite{Garavito-Camargo2021b}.
Recently, the instantaneous velocity offset of the inner MW with respect
to the outer halo was directly measured using tracer stars in the stellar halo
of the MW (the ``travel velocity''; \citealt{Petersen2021}).
Thus, the newly measured travel velocity can be used in Timing Argument studies
in place of orbital modeling to account for the expected perturbations of the
inner MW halo without having to make assumptions about the mass or dynamical
history of the LMC or other satellites.

Studies have considered variations to the standard Timing Argument model.
For example, one such model considered the effect of dark energy, and finds that
the addition of a cosmological constant to the energy equations yields a
$\sim13\%$ increase in the recovered mass~\citep{Partridge:2013}.
Similarly, the travel velocity of the MW disk, which has only recently been
first measured, introduces its own complication to the standard Timing Argument
model.

In this Article, we quantify the impact of this newly measured MW disk
motion on LG mass measurements using the Timing Argument,
thus accounting for observational
misinterpretations in a model-independent way for the first time.
We also show that improvements in the measurements of the MW travel
velocity may lead to even larger discrepancies between TA schemes with and
without a MW travel velocity.
We also explore a combination of recent measurements of the distance and proper
motions of M31 to infer the effect of the travel velocity in a
data-set-independent way.
As a result, we find that the travel velocity significantly impacts the inferred
mass of the LG in Timing Argument studies, and thus must be accounted for in
further dynamical studies of the LV.

\section{Methods and Data}
\subsection{Dynamical Model: The Timing Argument}
\label{sec:timingarg}
Following past work that utilizes the ``Timing Argument,'' we assume that the
orbital trajectories of the MW and M31 --- the Local Group system --- over
cosmic history are well described by Keplerian orbits \citep[e.g.,][]{Kahn1959,
Lynden-Bell:1981, Kroeker1991, LiWhite2008, vdm2012, Penarrubia2016}.
By assuming that M31 and the MW are gravitationally bound and were last at
closest approach in the early universe (i.e., the two galaxies have not yet
strongly interacted), we can then use the present-day kinematics of M31 relative
to the MW to estimate the total mass of the LG (i.e., using the
Timing Argument).

In this work, we largely follow the methodology and notation defined in
\citet{Penarrubia2016}.
Briefly recapping the classical Timing Argument method, we assume that the
dynamics of the
MW and M31 pair is dominated by the local gravitational potential of the LG, and therefore the Hubble flow can be neglected for computing the
relative orbits of the galaxies \citep[see, e.g.,][]{Penarrubia2014}.
Since we observe the relative position and motion between M31 and the MW, we
reduce the dynamics of the galaxies in the LG system to a single Keplerian orbit
that specifies the relative orbit between the galaxies and is completely
determined by four model parameters: the total mass of the LG, \mlg, the
semimajor axis, $a$, the eccentricity, $e$, and the present value of the
eccentric anomaly, $\eta$.

In terms of these four model parameters, the closed-form equations for relevant
two-body quantities that are closer to observables, like the separation between
the masses, $r$, the elapsed time since last pericenter, $t$, and the radial and
tangential velocity components, \vrad\ and \vtan, are given by
\begin{align}
  r &= a \, (1-e\,\cos\eta) \label{eq:r} \\
  t &= \left( \frac{a^3}{GM} \right)^{1/2}(\eta-e\,\sin\eta) \label{eq:t} \\
  \vrad &= \left( \frac{GM}{a} \right)^{1/2} \frac{e\,\sin\eta}{1-e\,\cos\eta}
  \label{eq:vrad} \\
  \vtan &= \left( \frac{GM}{a} \right)^{1/2} \frac{\sqrt{1-e^2}}{1-e\,\cos\eta}
  \label{eq:vtan} \quad .
\end{align}
In the expressions above, $r$ is the separation between the centers of the MW
and M31 halos, the time since last pericenter, $t$, is the age of the Universe,
and the velocity components, $(\vrad, \vtan)$, express the radial and tangential
velocity components of M31 relative to the center of the MW halo.

In a simpler universe where the MW and M31 are point masses and there are no
other massive bodies in the LG system, we could transform the observed
heliocentric sky position, distance, and velocity of M31 to a MW Galactocentric
reference frame and combine these with an estimate of the Hubble time to obtain
the four ``observables'' $(r, t, \vrad, \vtan)$.
These observables would be enough to infer the four model parameters $(\mlg, a,
e, \eta)$ using Equations~\ref{eq:r}--\ref{eq:vtan}.

To describe this ``classical'' Timing Argument approach in more detail and set
the stage for extending it, we adopt the notation of \citet{Penarrubia2016} in
which $\vel{A}{B}$ represents the velocity vector of A as measured in
the reference frame of B and $\pos{A}{B}$ represents the position vector of A as
measured from B.
With this notation, $\vel{A}{B} = -\vel{B}{A}$ and $\vel{A}{C} = \vel{A}{B} +
\vel{B}{C}$.

In the classical Timing Argument, the MW disk and M31 are assumed to occupy the
center of the potential well of their dark matter halos and have zero velocity
with respect to the halos.
We refer to this reference frame in the MW dark matter halo
as `\mwouter'.
Thus, the position and velocity of M31 with respect to the MW can be
represented by $\pos{M31}{\mwouter}$ and $\vel{M31}{\mwouter}$, which are
assumed to be equivalent to the position and velocity of M31 with respect to the
center of the MW disk.
Then, the \textit{observed} position and velocity of M31, measured in a heliocentric
reference frame, are given by
\begin{align}
  \pos{M31}{\odot} &= \pos{M31}{\mwouter} + \pos{\mwouter}{\odot}
  \label{eq:xoffset1}\\
  \vel{M31}{\odot} &= \vel{M31}{\mwouter} + \vel{\mwouter}{\odot}
  \label{eq:voffset1}\quad .
\end{align}
Here $\left|\pos{M31}{\mwouter}\right| = r$ as determined from
Equation~\ref{eq:r}, $\vel{M31}{\mwouter}$ is determined completely by the
Keplerian model parameters (through \vrad\ and \vtan), and the position and
velocity of the center of the MW halo as measured from the sun are
$\pos{\mwouter}{\odot}$ and $\vel{\mwouter}{\odot}$.

However, the true dynamics of the MW--M31 system are not so simple.
Perturbations introduced by interactions and mergers between the MW and its
satellite galaxies, such as the merger of the Sagittarius dwarf galaxy or the
recent pericentric passage of the LMC, break the assumption that the MW disk
is stationary in the center of its dark matter halo.
In fact, these interactions will introduce an additional reflex motion component
in observations from the MW disk compared to the \mwouter\ reference frame.
The LMC's impact on the dynamics of the MW disk and inner halo have been studied
in detail by~\cite[e.g.,][]{Gomez2015, Garavito-Camargo:2019, Petersen:2020,
Garavito-Camargo2021b}.
These works imply that we must include additional terms in
Equations~\ref{eq:xoffset1} and \ref{eq:voffset1} to account for the travel
velocity of the MW disk with respect to the center of the halo in its
unperturbed state.
Thus, the observed position and velocity vectors of M31 from the solar reference
frame become
\begin{align}
  \pos{M31}{\odot} &= \pos{M31}{\mwouter} + \pos{\mwouter}{\mwdisk}+\pos{\mwdisk}{\odot} \label{eq:xm31-solar}\\
  \vel{M31}{\odot} &= \vel{M31}{\mwouter} + \vel{\mwouter}{\mwdisk}+\vel{\mwdisk}{\odot}\label{eq:vm31-solar}
\end{align}
where ``\mwouter'' refers to a reference frame centered at and moving with the
center of mass of the outer MW halo, ``\mwdisk'' refers to a reference frame
centered at and moving with the center of the MW disk,
and $\pos{\mwdisk}{\odot}$ and $\vel{\mwdisk}{\odot}$, respectively, are
the adopted solar position and velocity in the Galaxy. The values we
adopt for $\pos{\mwdisk}{\odot}$ and $\vel{\mwdisk}{\odot}$
(shortened to $\boldx_{\odot}$ and $\bov_{\odot}$)
are given in Table~\ref{table:data} below.\footnote{
Note that in principle, there is also a term $\vel{\rm M31_{halo}}{\rm
M31_{disk}}$; however, there are not yet measurements of the differential
motion of the M31 disk with respect to the M31 halo, so we neglect this term.
See further discussion in Section~\ref{sec:discussion-m31}.
}

Observationally, the reflex motion of the disk imprints itself on velocity
measurements as an instantaneous velocity shift.
Recently,~\cite{Petersen2021} used tracers in the outer MW stellar halo to
measure this instantaneous travel velocity
$\vtrav=\vel{\mwouter}{\mwdisk}$.
They find $\big|\vtrav\big|=32\pm4\kms$ with a highest likelihood apex
direction in Galactocentric coordinates of
$(\ell, b)_{\rm apex} = (56^{+9}_{-9},-34^{+10}_{-9})$ degrees.
We assume that $\pos{\mwouter}{\mwdisk} \approx 0$ motivated by the fact
that this displacement is likely much smaller than the distance between the MW
and M31 $\pos{\mwouter}{\mwdisk} \ll r$ \citep[as expected from simulations;
e.g.,][]{Garavito-Camargo2021b}.
However, it is important to note that this displacement is still significant on
scales relevant for many other MW studies
(see Section~\ref{sec:discussion-impact} for more details).

\begin{figure*}[htb]
  \centering
  \includegraphics[width=0.8\textwidth]{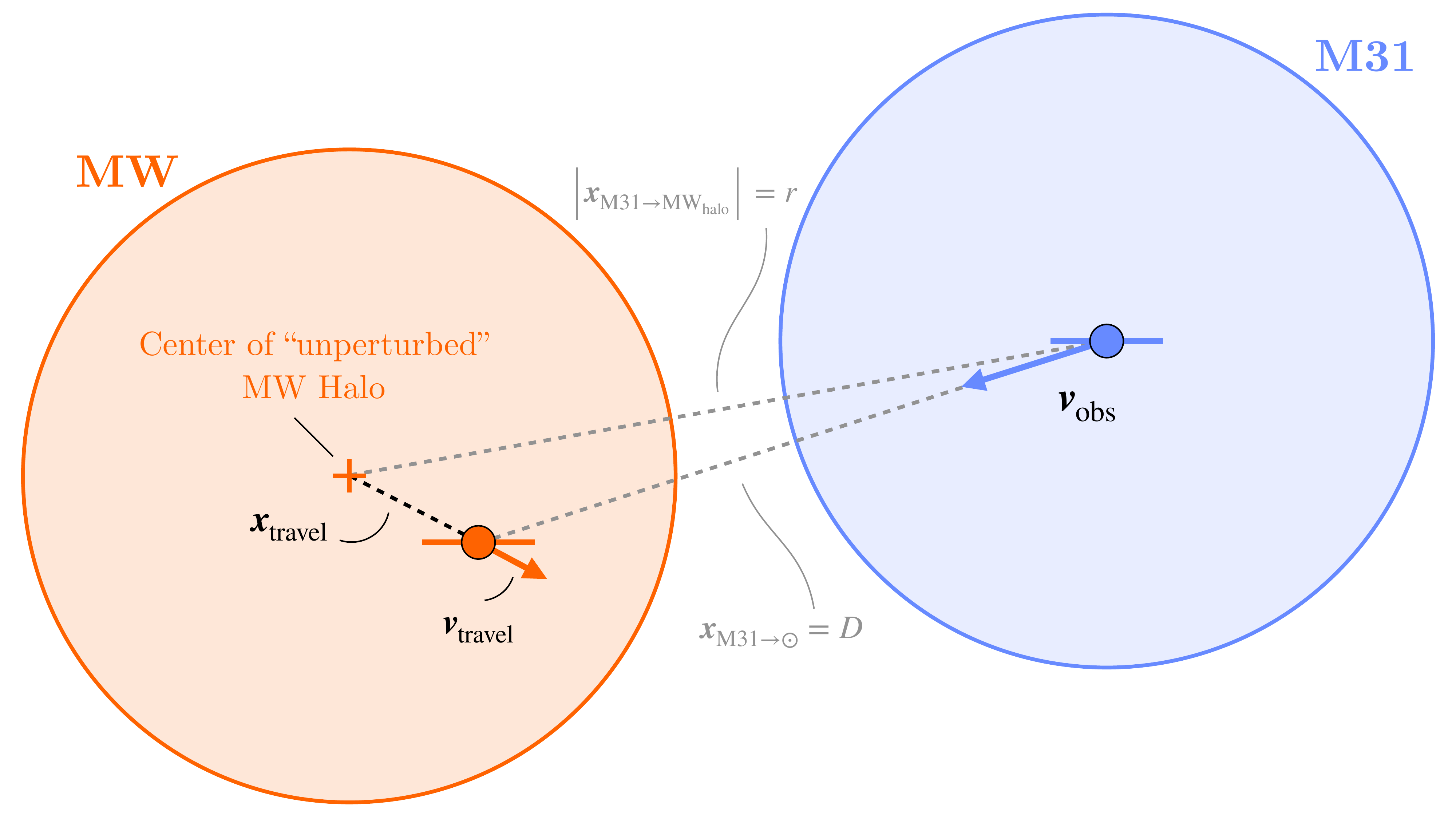}
  \caption{
    Schematic of the Milky Way (MW) and M31 system, not to scale. The shaded regions
    represent the halos of both galaxies. Shown are an artistic representation
    of the velocity and position vectors that are relevant in our model,
    including
    $D=\pos{M31}{\odot}$, the measured distance to M31;
    $\pos{M31}{\mwouter}$, the distance between the centers of both halos;
    and $\boldsymbol{v}_{\rm obs}$, the measured 3D velocity of M31.
    Finally, \xtrav\ and \vtrav\ are the present distance and velocity
    between the center of the MW halo and the center of the MW disk.
    For this study, we assume $\xtrav\ll r,D$, and $|\vtrav| = 32\pm4 \kms$
    from~\cite{Petersen2021}.
    }
  \label{fig:schematic}
\end{figure*}

Figure~\ref{fig:schematic} shows a schematic of these different vectors --- all
drawn in a frame that is comoving with the \mwouter\ frame --- and a rough
illustration of the geometry we assume. For clarity, we show
$\pos{\mwouter}{\mwdisk}= \xtrav$, $\vel{\mwouter}{\mwdisk}=\vtrav$, and
$\vel{M31}{\odot}=\bov_{\rm obs}$.

\subsection{Data Sets}
\label{sec:data sets}
The present distance and relative velocity of M31, as well as the age of the
universe (used in Equation~\ref{eq:t}), are key observables that are used to
constrain our Timing Argument model. In this paper, we consider three different
compilations of data to understand how different measurements might affect the
Timing Argument model with the addition of the travel velocity. In particular,
we consider two different M31 distance measures: an approximated distance
measure from~\cite{vdm2008}, and a more accurate Cepheid-based distance measure
from~\cite{Li2021}. We also consider two different M31 proper-motion
measurements: Hubble Space Telescope (\hst)-based proper motions from~\cite{vdm2012}, and a more recent
\gaia\ early Data Release 3 (eDR3)-based proper-motion measure from~\cite{Salomon2021}.

We have split these measurements into three compiled data sets:
\begin{description}
  \item[vdMG08 Dist. + HST PM:] the
  M31 distance measure from~\cite{vdm2008} and \hst\ proper motions
  from~\citet{vdm2012}, the same data set as used to constrain the LG
  mass via the Timing Argument
  in~\citet{vdm2012}
  \item[Cepheid Dist. + Gaia PM:]  a
  compilation of more recent M31 kinematic measurements, including a more
  precise Cepheid-based distance measure to M31 \citep{Li2021} and updated
  \gaia\ eDR3 proper motions from \citep{Salomon2021}.
  \item[Cepheid Dist. + HST PM:] a hybrid data set with the Cepheid-based distance
  measure to M31 and the \hst-based proper-motion measurement
  \citep{Li2021,vdm2012}.
\end{description}
The \hst-based proper motions were originally presented in~\cite{Sohn:2012},
and were then corrected for the internal kinematics and space motion of
M31 in~\cite{vdm2012},
from which we used the ``Weighted Average'' heliocentric
velocities in Table~3.\footnote{
Note that the referenced papers actually report velocity components and
uncertainties: To transform from velocity back to proper motions, we divide out
the adopted distance and deconvolve the distance uncertainty to obtain proper
motions and uncertainties of $\mu_{\alpha^*} = 34.30 \pm 8.25~\masyr$ and
$\mu_{\delta} = -20.22 \pm 7.71~\masyr$.
}
The \textit{Gaia}-based M31 proper-motion measurement is slightly larger than
the \hst\ proper motion of M31, leading to an increased implied transverse
velocity that, a priori, should lead to a higher inferred LG mass
compared to the more radial orbit implied by the \hst\ proper motions.
See Table~\ref{table:data} for numerical values used in each of these data sets.

\begin{table*}
  \begin{tabular}{lc|c|c}
    \hline\hline
      & \textbf{vdMG08 Dist. + HST PM} & \textbf{Cepheid Dist. + Gaia PM} & \textbf{Cepheid Dist. + HST PM}\\\hline
  $D~[\kpc]$ & $770 \pm 40\reflabel{a}$ &   $761\pm11~\kpc\reflabel{f}$  & $761\pm11\reflabel{f}$\\
  $v_{\rm rad}~[\kms]$ & $-301 \pm 1\reflabel{b}$ & $-301\pm 1\reflabel{b}$ & $-301\pm 1\reflabel{b}$ \\
  $\mu_{\alpha^*}~[\muasyr]$    & $34.30\pm 8.25\reflabel{c}$  & $48.98\pm 10.47\reflabel{g}$ & $34.30\pm 8.25\reflabel{c}$ \\
  $\mu_\delta~[\muasyr]$ & $-20.22 \pm 7.71$\reflabel{c} & $-36.85\pm 8.03\reflabel{g}$ & $-20.22 \pm 7.71$\reflabel{c} \\
  $\bs{x}_\odot$~[\kpc]& $(-8.29, 0, 0)\reflabel{d}$ & $(-8.122, 0, 20.8)\reflabel{h}$ & $(-8.122, 0, 20.8)\reflabel{h}$ \\
  $\bs{v}_\odot$~[\kms]& $(11.1, 251.54, 7.25)\reflabel{d}$ & $(12.9, 245.6, 7.78)\reflabel{i}$ & $(12.9, 245.6, 7.78)\reflabel{i}$ \\
  $t_{\rm peri}~[\Gyr]$ & $13.75\pm 0.11\reflabel{e}$  & $13.801 \pm 0.024$ \reflabel{j} & $13.801 \pm 0.024$ \reflabel{j}\\
  \hline\hline
  \end{tabular}
  \tablerefs{$a.$ \cite{vdm2008},
   $b.$ \cite{Courteau1999},
   $c.$ \cite{vdm2012},\\
   $d.$ \cite{Schonrich2010,McMillan2011},
   $e.$ \cite{Jarosik2011},
   $f.$ \cite{Li2021},
   $g.$ \cite{Salomon2021},\\
   $h.$ \cite{GravityCollab2018,Bennett2019},
   $i.$ \cite{Drimmel2018},
   $j.$ \cite{Planck2018}}
  \caption{\label{table:data}
  Observational data sets used for comparison throughout analysis and their
  references. Each value is measured for M31 with
  respect to the sun. $D$ is the distance, $v_{\rm rad}$ is the radial velocity,
  and $(\mu^*_{\alpha}, \mu_{\delta})$ are proper motions in RA cosdec and Decl.
  $\bs{x}_\odot=(x, y, z)$ and
  $\bov_{\odot}=(\rm U_{\rm pec}, V_{\rm pec}+V_0, W_{\rm pec})$ are the
  the position of the Sun and the solar motion with respect to the Galactic
  center, with the x-axis pointing from the projection of the Sun
  on the disk towards the Galactic center, and the z-axis pointing in the
  direction of the North Galactic Pole.
  $t_{\rm peri}$ is the time elapsed since the last pericenter of the M31
  Keplerian orbit, which in this case is the age of the Universe.
  }
\end{table*}

\subsection{Bayesian Inference}
\label{sec:bayes}

We construct a likelihood function, $\mathcal{L}$, to quantify the probability of
measuring the observed quantities $\vec y = \{D, \vrad, \mu^*_{\alpha}, \mu_{\delta},
t_{\rm peri}\}$ given a Timing Argument model with parameters $\vec \theta = (\mlg, a,
e, \eta, \alpha)$.
Here, $D$ the distance to M31, $\vrad$ the radial velocity of M31, $\mu^*_{\alpha}$ and
$\mu_{\delta}$ the proper-motion components, and $t_{\rm peri}$ the time since last
pericenter.
The parameter vector $\vec \theta$ contains $\mlg$ the total mass of the LG, $a$ the
semimajor-axis, $e$ the eccentricity, $\eta$ the present value of the eccentric anomaly,
and $\alpha$ a nuisance parameter discussed in detail later in this section.
We assume that the measurements are independent and have Gaussian uncertainties such
that the likelihood function is a product:
\begin{equation}
  \mathcal{L} = p(\vec y|\vec \theta)\\=
  \prod^n \frac{1}{\sqrt{2\pi} \sigma_{n} } \exp\left[ -\frac{1}{2}\,\frac{(y_n - \tilde{y}_n(\vec{\theta}))^2}{2\sigma_n^2} \right]
\end{equation}
where $n$ indexes the elements of the data vector, $\sigma_{n}$ is the corresponding
uncertainty for the $n$th data element, and $\tilde{y}_n(\vec{\theta})$ is the
model-predicted value for a given data component.

We then adopt prior probability distribution functions (pdfs) for the parameters
and use these pdfs to compute the posterior pdf over the parameters given the
data in order to generate samples from the posterior pdf using a Markov Chain
Monte Carlo (MCMC) method.
In order to recover the estimated mass distribution, we marginalize over
all other model parameters.

In detail, we first use the four Timing Argument parameters to compute the
present-day separation between the MW and M31 halos and their relative radial
and tangential velocities as defined in Equations~\ref{eq:r}--\ref{eq:vtan}.
These velocity components represent the relative velocity M31 would have as
observed from the center of an unperturbed MW halo.
We then use the measured ``travel velocity'' of the MW disk,
\vtrav, to find the relative velocity of M31 with respect to the center of the
moving MW disk (i.e., a moving MW Galactocentric frame).
We finally transform from this Galactocentric frame to a heliocentric
reference frame moving with the solar system barycenter (i.e., ICRS coordinates).
At this final stage, we must introduce an additional nuisance parameter $\alpha$
that represents the orientation of the MW--M31 orbital plane as it intersects
the tangent plane located at the sky position of M31 as viewed from the MW disk
center. This parameter is needed to convert from the two-dimensional velocity
components given by Equations~\ref{eq:vrad}--\ref{eq:vtan} to the
three-dimensional velocity components represented by the two proper-motion
components and the radial velocity of M31.
However, we stress that this position angle has no impact on the fundamental
dynamical parameters and is only used for coordinate transformations.

We specify this model using the \texttt{Python} probabilistic programming
package \texttt{pymc3}~\citep{Salvatier2016} and use the No-U-Turn Sampler
(NUTS) \citep{Homan2014} implemented in \texttt{pymc3} to generate samples from
this posterior pdf, given data from each of the data sets defined in
Table~\ref{table:data}.
We sample over the parameters LG mass $\mlg$, the present-day MW--M31
halo separation $r$, log eccentricity $\ln\left(1 - e\right)$, eccentric anomaly
$\eta$, and the orbital plane orientation nuisance parameter $\alpha$.
Our adopted prior pdfs are defined in Table~\ref{table:priors}.
For each data set, we run the sampler with four chains for 4000 tuning steps and
40,000 draws.

\begin{table}
  \centering
  \begin{tabular}{lc}
  \hline\hline
  Prior  & Description \\\hline
  \mlg: $\mathcal{N}_T(4.5,3)\times10^{12}\Msun$ & Mass of the Local Group\\
  $r$: $\mathcal{N}_T(700,100)$kpc & Distance from M31 to $\mwdisk$\\
  $\ln(1-e)$: $\mathcal{U}(-10,0)$ & Eccentricity (close to 1) \\
  $\eta$: $\mathcal{U}(-\pi, \pi)$ & Eccentric anomaly\\
  \multirow{2}{*}{$\alpha$: $\mathcal{U}(-\pi, \pi)$} & Position angle of M31
  orbital\\
  & plane from MW disk center\\
  \hline\hline
  \end{tabular}
  \caption{\label{table:priors} A description of our adopted prior probability
  distribution functions over the Timing Argument model parameters.
  Here, $\mathcal{U}(a, b)$ represents a uniform distribution over the domain
  $(a, b)$, and $\mathcal{N}_T(\mu, \sigma)$ represents a truncated Normal
  distribution with mean $\mu$ and standard-deviation $\sigma$.
  We truncate the mass prior pdf to the range $(0.5, 20)\times10^{12} \Msun$ and
  the distance prior pdf to the range $(100, 10^{4})\,\kpc$.
  }
\end{table}

\section{Results: Local group mass estimates}
\label{sec:results}
\begin{figure}[htb]
  \centering
  \includegraphics[width=0.5\textwidth]
  {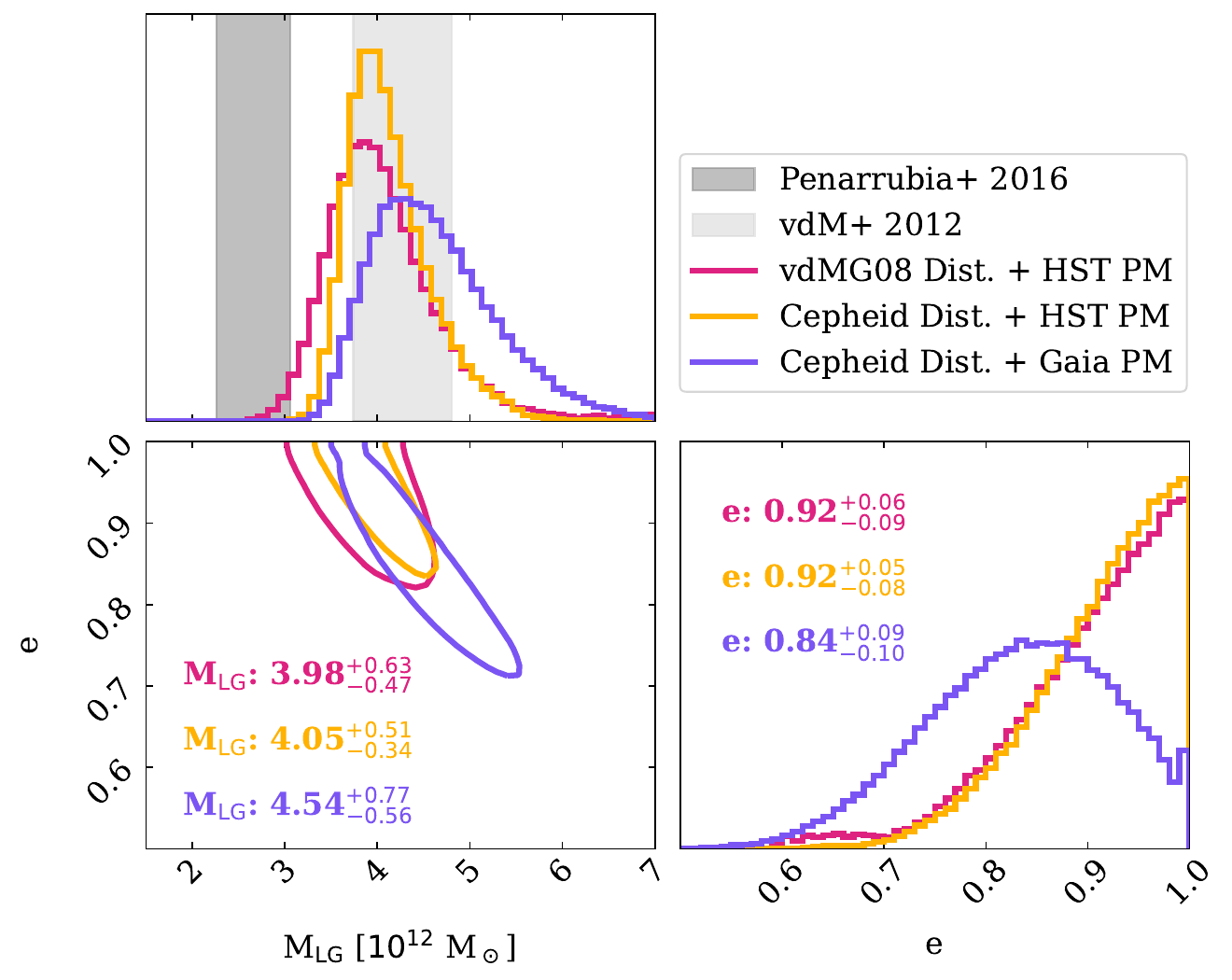}
  \caption{\label{fig:contour} Sixty-eight percent credible regions of sampled posterior
  distributions with three observational data sets for a subset of our model
  parameters: the total mass of the Local Group (\mlg)
  and the eccentricity of the orbit of M31 about a fixed MW ($e$).
  Mean masses (in units of $10^{12}\Msun$) and eccentricities are reported in
  the bottom left and right panels along with the 68\% credible region for each
  data set.
  The shaded regions in the upper left panel are the 68\% credible region mass
  estimates of previous TA studies from~\cite{vdm2012}
  and~\cite{Penarrubia2016}.
  The more radial orbit implied by the~\cite{vdm2012} \textit{HST} proper
  motions leads to a lower inferred \mlg\ and a higher
  eccentricity, while the larger \textit{Gaia} proper motions of
  ~\cite{Salomon2021} yield a more circular orbit, and thus a lower eccentricity
   and higher mass system.
   }
\end{figure}

We use a Bayesian implementation of a Timing Argument model to quantify the
impact of the measured travel velocity of the MW disk from~\cite{Petersen2021}
on the estimated mass of the LG, as well as other orbital parameters
such as the distance between M31 and the MW.
We compute convergence statistics using \texttt{Arviz}~\citep{arviz} for all
MCMC runs and find that the maximum Gelman--Rubin convergence statistic is
$\leq1.01$ for all parameters and each data set~\citep{GelmanRubin1992}.
The mean inferred parameter values and their 68\% credible regions from the
sampled posterior pdfs for each data set are presented in
Table~\ref{table:convergedparams}.
These results are also shown in
Figure~\ref{fig:contour}, again displaying the 68\% credible regions of the LG
mass, \mlg\, and eccentricity, $e$ (lower left panel).
The upper-left panel shows the marginal posterior pdfs over LG mass for each of
the data sets (histogram curves) and with 68\% credible regions plotted for two
prior LG mass measurements (gray shaded bands).
We note that these values are for reference only, as we currently do not account
for other known sources of bias in the TA model -- such as a cosmological constant or
cosmic bias -- that may yield more accurate values.

We find that the addition of the travel velocity of the MW disk systematically decreases
the inferred LG mass and eccentricity of the
orbit compared to models that do not include the travel velocity of the MW disk.
We also find that the inferred mass is larger and the LG orbit is less eccentric
when using the (larger) \gaia\ proper-motion of M31.
For all data sets, the eccentricity of the decreases by $\sim~3-5\%$.
However, we find that the inferred orbital eccentricity is consistent with a
radial orbit.

\begin{figure}[htb]
    \centering
    \includegraphics[width=\columnwidth]{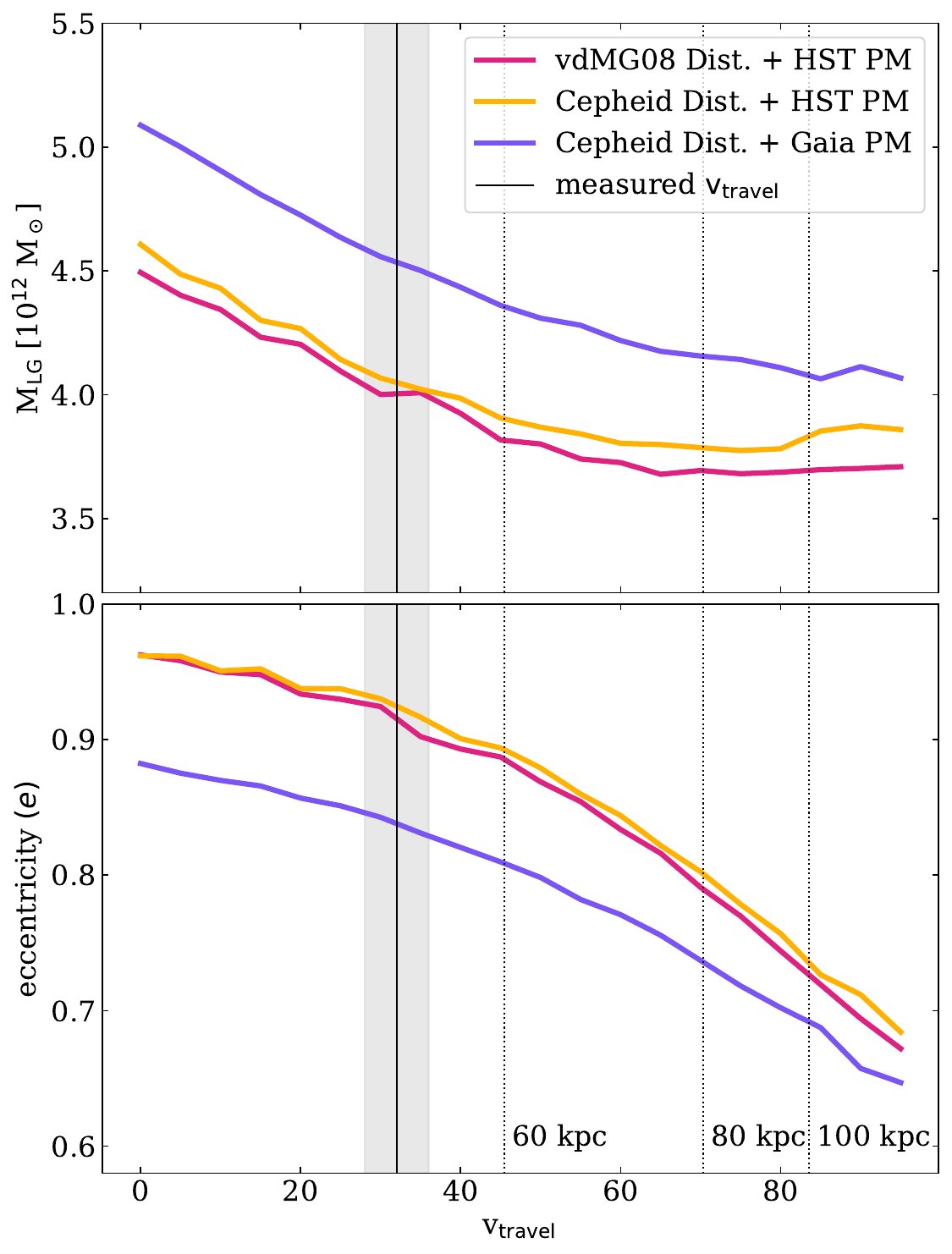}
    \caption{\label{fig:mvsv}
    Mean inferred Local Group (LG) mass (top) and eccentricity (bottom) as a function
    of travel velocity magnitude of the MW disk.
    The larger \textit{Gaia} proper motions (purple) lead to higher transverse
    motion and thus higher mass and a less eccentric orbit than either of the
    \textit{HST} proper motion data sets (pink and yellow), though each data set
    displays the same general trend with increasing travel velocity.
    The solid vertical line and accompanying shaded region represent the median
    and 67\% confidence interval of the travel velocity measured
    by~\cite{Petersen2021} of $\vtrav=32\pm 4\kms$.
    The dotted vertical lines represent simulated travel velocities for
    stellar tracers at different distances in~\cite{Garavito-Camargo2021b}.
    \textbf{The inclusion of the travel velocity of the MW disk systematically
    lowers the inferred mass and eccentricity of the Local Group} regardless of
    observational data set. A larger measured travel velocity will yield a lower
    mass, less radial Local Group.}
  \end{figure}

Figure~\ref{fig:mvsv} gives a summary of our key results, showing the
behavior of the inferred LG mass and eccentricity as a function of the travel
velocity. As the travel velocity increases from $\vtrav=0$ to $\vtrav=32\pm4$
km s$^{-1}$, shown by the vertical black line and gray shaded regions, the mass
and eccentricity both decrease, with the effect on the mass drastically changing
by up to $\sim0.5\times 10^{12}\Msun$ and the MW--M31 orbit becoming more
circular.

\begin{table*}
  \begin{tabular}{lc|c|c}
    \hline\hline
    Parameter  & \textbf{vdMG08 Dist. + HST PM} & \textbf{Cepheid Dist. + Gaia PM} & \textbf{Cepheid Dist. + HST PM}\\\hline
    $\mlg$ & $3.98^{+0.6}_{-0.5}$ & $4.54^{+0.8}_{-0.6} $ & $4.05^{+0.5}_{-0.3} $ \\
    $e$ & $0.92^{+0.1}_{-0.1}$ & $0.84^{+0.1}_{-0.1}$ & $0.92^{+0.1}_{-0.1} $ \\
    $r$ & $777.72^{+36.6}_{-36.0}$ & $765.17^{+10.9}_{-10.9}$ & $765.44^{+10.8}_{-10.9} $ \\
    $\eta$ & $-2.14^{+0.05}_{-0.04}$ & $-2.08^{+0.04}_{-0.04}$ & $-2.11^{+0.03}_{-0.03} $\\
    $\alpha$ & $2.97^{+0.8}_{-0.8}$ & $1.42^{+0.6}_{-0.5} $ &  $2.96^{+0.8}_{-0.8} $ \\
  \hline\hline
  \end{tabular}
  \caption{\label{table:convergedparams}Mean inferred parameter values and the
  68\% credible region of the sampled posterior region for each data set.
  Here, $\mlg$ is the mass of the Local Group, $e$ is the eccentricity of the
  MW--M31 orbit, $r$ is the distance between the centers of the MW and M31
  halos, $\eta$ is the eccentric anomaly (a proxy for the phase of the orbit),
  and $\alpha$ is a nuisance parameter representing the angle between the
  orbital plane of MW--M31 and the tangent plane located at the sky position of
  M31 as seen from the center of the MW disk.}
\end{table*}

\section{Discussion}
\label{sec:discussion}

\subsection{Comparing recent measurements of the Local Group mass}
There have been two primary pathways toward measuring the mass of the LG:
measure the masses of MW and M31 individually and add them, or go after the
total mass directly with LV dynamics, the Timing Argument, or cosmological
simulations.

Historically, the total mass estimates have been much larger than the sum of the
individual MW and M31 masses:
typical total LG masses are upwards of $4\times 10^{12}\Msun$, while the sum of
independent MW+M31 mass measures result in a total mass closer to
$2-2.5\times 10^{12}\Msun$, as can be seen in the collection of previous mass
estimates for the LG, M31, and the MW in Table~\ref{table:masses}.
It is not surprising that there are many discrepancies between the total and
summed values of the LG mass, since MW+M31 (i.e., individual summed) mass
estimates often cannot measure the full extent of the distribution of dark
matter within each galaxy, let alone in the LG, and thus must extrapolate in
regions where there may be large uncertainties.

There have been a few exceptions in the trend of high total mass measurements,
namely in~\cite{Diaz2014} and~\cite{Penarrubia2016}.
\cite{Diaz2014} utilize the fact that the LG momentum should balance to zero in
the frame of the LG barycenter to determine the total mass of the LG, as well as
the mass ratio between M31 and the MW. Using the LG barycenter, indicated by a
set of 17 LG satellites at $>350\kpc$ from the MW and M31, and the velocities of
M31 and the MW with respect to the barycenter, they found a LG mass of
$2.5\pm0.4\times10^{12}\Msun$, and a mass ratio $\rm M_{M31}/M_{MW}>2.29$.
The impact of the LMC and M33 are absorbed by assuming they contribute to the
masses of their host galaxies.
However, the recent measurement of the MW travel velocity will change the
measured barycenter and velocities of the MW, M31, and the LG satellites, though
it is unclear how this would affect the total mass in their analysis.

Additionally, ~\cite{Penarrubia2016} found a Local Group mass of
$\sim2.64\pm0.4\times 10^{12}~\Msun$, which is significantly lower than our
findings, although this constraint combines the Timing Argument dynamics of
the M31--MW system in addition to the observed kinematics of 35 LV
galaxies. They parameterize the offset of the LMC+MW barycenter from a
MW-only barycenter as a function of the mass ratio between the LMC and MW, and
find that the LMC likely has~$\sim25\%$ of the mass of the MW halo, resulting in
a large shift in the barycenter of the LMC+MW.
The recovered mass using only the dynamics of the LV galaxies was quite low
($\sim 2\times 10^{12}\Msun$), though
including the Timing Argument dynamics of the MW--M31 system increased their
recovered total mass to $\sim2.64\pm0.4\times 10^{12}~\Msun$.

Our TA$+$\vtrav\  LG mass estimates are consistent with a number of other recent
studies that estimate the mass of the LG through dynamical methods.
For example, we find agreement with the previous Timing Argument model
of~\cite{vdm2012} which found a total mass
of $4.27\pm0.45\times 10^{12}\Msun$ (neglecting for cosmic bias and
scatter) using the same values for the distance and velocity of
M31 as in our \textit{vdMG08 Dist. + HST PM} data set.

In two $N$-body cosmological simulations, Millenium-WMAP7 and MilleniumII,
\cite{Zhai2020} identified pairs of stellar analogs to the MW and M31, then
applied a series of kinematic cuts on the separation, isolation, and velocities
of the pair to determine LG analogs.
They find stellar and dynamical LG analogs on mostly radial orbits have total
masses of $4.4^{+2.4}_{-1.5}\times 10^{12}\Msun$, which is consistent with our
findings for each data set.
They also find that low-ellipticity orbits (where
$v_{\rm rad}\sim v_{\rm tan}$), result in a higher LG mass, M31 mass, and MW
mass, reporting $6.6^{+2.7}_{-1.5}\times10^{12}\Msun$,
$3.8^{+2.8}_{-1.8}\times10^{12}\Msun$, and
$2.5^{+2.2}_{-1.4}\times10^{12}\Msun$, respectively.
We find that the ellipticity of the orbit decreases dramatically with larger
travel velocities; however, in contradiction to \cite{Zhai2020}, we find that
this leads to a lower total LG mass.

Additionally, \cite{Benisty2022} recently modeled the contribution of the
LMC-induced shift in the barycenter of the MW and used the Timing Argument to
place constraints on the LG mass while removing the impact of the LMC.
More specifically, they estimate the contribution of the reflex motion of the
MW disk to the observed velocity vector of M31 by modeling the orbital history
of the MW--M31 system and the MW+LMC--M31 system.
The impact of the LMC on the mass measurements of the LG thus depend on the
orbital and mass models of the LMC about the MW.
This is in contrast to our work, which need make no assumption about the mass of
the LMC, its orbital history, or the merger history of the MW.
Upon implementing the Timing Argument, including a cosmological constant,
and removing the LMC-induced reflex motion,
they find a LG mass of $5.6^{+1.6}_{-1.2}\times10^{12}\Msun$, which is roughly
25\% larger than our findings, and that accounting for cosmic bias \& scatter
lowers the mass by an additional 40\% to $3.4^{+1.4}_{-1.1}\times10^{12}\Msun$,
25\% lower than our findings.
However, they find that, in general, inclusion of the motion of the MW disk due
to the interaction with the LMC lowers the LG mass by $\sim10\%$, which is
consistent with our finding of a reduction in the total LG mass by $\sim10-12\%$.

A notable difference between this study and~\cite{Benisty2022} is the method by
which the reflex motion of the MW disk is accounted for.
Rather than relying on accurately simulating the reflex motion of the disk, we
let the travel velocity of the MW disk introduce a coordinate transformation
(boost) of the measured velocity vectors of M31, and fit for the model
parameters given the observable data.
This method allows us to avoid model uncertainties in the mass profiles of each
galaxy in the orbital models of the interaction between the MW, M31, and the LMC.
Additionally, using the measured travel velocity also allows us to innately
account for possible additional contributions to the present-day, instantaneous
travel velocity induced by the extensive merger history of the MW in a robust
way, without the need to simulate the entire interaction history.

\begin{table*}
  \centering
  \begin{tabular}{clc|c}
    \hline\hline
    Mass & Method & Result ($ 10^{12}~\Msun$ ) & Citation \\\hline
    \multirow{15}{*}{\mlg}  & TA & $3.6$ & \citealt{Lynden-Bell:1981} \\
    &TA (radial + cosmo sim calibration)  & $5.27$& \citealt{LiWhite2008} \\
    &{TA only} & {4.27$\pm$0.45} & \citealt{vdm2012} \\
    &{TA (3D + cosmic bias and scatter)} & {4.93$\pm$1.63} &\citealt{vdm2012}\\
    & LG Dynamics &$2.5\pm0.4$ & \citealt{Diaz2014}\\
    &{LV Galaxies + TA + $\Lambda$} & {2.64$\pm0.4$} & \citealt{Penarrubia2016} \\
    & Machine Learning & $4.9 \pm 0.8$ & \citealt{McLeod2017}\\
    & Machine Learning (+large M31 transverse motion) & $3.6\pm0.3$ &
    \citealt{McLeod2017}\\
    & Cosmological Sims & 4.4$^{+2.4}_{-1.5}$ & \citealt{Zhai2020}\\
    & Cosmological Sims (likelihood-free inference) & $4.6^{+2.3}_{-1.8}$ &
    \citealt{Lemos2021}\\
    & TA + $\Lambda$ + Cosmological Sims & $4.75^{+2.22}_{-2.41}$ & \citealt{Hartl2022}\\
    & TA + $\Lambda$ + LMC & 5.6$^{+1.6}_{-1.2}$ & \citealt{Benisty2022}\\
    & TA + $\Lambda$ + cosmic biad + LMC & 3.4$^{+1.4}_{-1.1}$ & \citealt{Benisty2022}\\
    & \textbf{TA + $\boldsymbol{\vtrav}$} (Cepheid + HST) & {$\boldsymbol{4.0^{+0.5}_{-0.3}}$}&
    \textbf{Chamberlain et al. 2022 (this work)}\\
    & \textbf{TA + $\boldsymbol{\vtrav}$} (Cepheid + Gaia) & {$\boldsymbol{4.5^{+0.8}_{-0.6}}$}&
    \textbf{Chamberlain et al. 2022 (this work)}\\
    \hline
    \multirow{9}{*}{\mmto}& Kinematics of M31 Sats & $1.4 \pm 0.4$ ($<$300 kpc)
    & \citealt{Watkins2010}\\
    & Giant Stellar Stream & $2.00^{+0.52}_{-0.41}$ & \citealt{Fardal2013}\\
    & LG Dynamics &$1.7\pm0.3$ & \citealt{Diaz2014}\\
    & Local Hubble Flow &  $1.33\pm0.4$ & \citealt{Penarrubia2016}\\
    & M31 Orbital Ang. Mom. & $1.37^{+1.39}_{-0.75}$ & \citealt{Patel2017b}\\
    & Cosmological Sims & $1.0-2.0$ & \citealt{Carlesi2017} \\
    & Cosmological Sims & 2.5$^{+1.3}_{-1.1}$ & \citealt{Zhai2020}\\
    & Machine Learning & $2.3-2.5$& \citealt{Villanueva-Domingo2021}\\
    & Machine Learning (+velocity information) & $2.2-2.5$ &
    \citealt{Villanueva-Domingo2021}\\

    \hline
    \multirow{9}{*}{\mmw}&  Kinematics of LG sats & $1.4 \pm 0.3$ ($<$300 kpc)
    & \citealt{Watkins2010}\\
    & LG Dynamics &$0.8\pm0.5$ & \citealt{Diaz2014}\\
    & Local Hubble Flow & $1.04 \pm 0.26$ & \citealt{Penarrubia2016}\\
    & LMC Orbital Ang. Mom. & $1.02^{+0.77}_{-0.55}$ & \citealt{Patel2017b}\\
    & Cosmological Sims & $0.6-0.8$ & \citealt{Carlesi2017}\\
    & MW Sats & $0.96^{+0.29}_{-0.28}$ & \citealt{Patel2018}\\
    & Cosmological Sims & 1.5$^{+1.4}_{-0.7}$ & \citealt{Zhai2020} \\
    & Machine Learning & $1.0-1.3$ & \citealt{Villanueva-Domingo2021}\\
    & Machine Learning (+velocity information) & $2.3-2.6$ &
    \citealt{Villanueva-Domingo2021}\\
  \hline\hline
  \end{tabular}
  \caption{\label{table:masses}A collection of previous mass
  measurements from LG dynamics discussed in the text focusing on previous Timing Argument results.}
\end{table*}

\subsection{Additional sources of bias to the Timing Argument}
Given the simplicity of the Timing Argument dynamical model --- in particular,
the assumption that the MW and M31 are point masses with constant masses --- it
is reasonable to wonder whether this methodology provides unbiased estimates of
the true LG mass.
An early study of a dark-matter-only cosmological simulation found that the
Timing Argument applied to pairs of galaxies did provide unbiased estimates of
the sum of masses of the pairs \citep{LiWhite2008}.
However, more recently it was found that conditioning on LG analogs with similar
radial and tangential velocities to the MW and M31 leads to slightly biased
(overestimated) inferred total masses of those systems \citep{Gonzalez2014,
Hartl2022}.
In this work, we do not attempt to ``correct'' our inferred LG masses for this
cosmic bias effect, because it is unclear whether cosmological simulations
accurately reproduce the detailed properties of LG systems.
Accounting for this effect would likely lower our reported LG mass measurements.
However, as we have shown, the existence of a reflex motion of the
MW disk as a response to the MW's interaction with its satellite population
decreases is an additional perturbation to the TA that must be considered in
future studies alongside cosmic bias and a cosmological constant.

\subsection{Impact of travel velocity on inferred dynamics of the Local Group}
\label{sec:discussion-impact}
The existence of the travel velocity of the MW disk
results in measurable differences in the estimated mass of the LG
through the Timing argument.
These results are consistent with \cite{Erkal2020}, who find that neglecting the
LMC-induced reflex motion of the MW can result in masses that are overestimated
by up to ~50\%.
As shown in our results above, neglecting this motion at its currently measured
value of $\vtrav=32\pm4\kms$
\citep{Petersen2021} leads to LG masses that are overestimated by $\sim30\%$.
However, both the magnitude and \emph{direction} of the travel velocity are
directly tied to the inferred mass of the LG.
As it is currently measured, the (highest likelihood) direction of $\vtrav$
is $\sim$$60^\circ$ from the sky position of M31, meaning that the travel
velocity impacts the conversion of both M31's observed proper motion and radial
velocity from a Heliocentric reference frame to the ``outer halo'' reference
frame used above.
At fixed magnitude, if the true apex of the travel velocity motion is closer
(farther) to M31's sky position, it would primarily affect the radial velocity
(proper motions).

There is reason to believe that the recently measured MW disk travel velocity
could be a lower bound on the true value, which could be up to a factor of
$\sim$2--3 higher than the currently measured value.
Using an idealized simulation of an equilibrium dark matter halo that has a
recent merger with an LMC-like halo, \cite{Garavito-Camargo2021b} showed that
stellar halo tracers at different distances from the MW disk center may result
in different measured travel velocities.
While this simulation does not span previous mergers in the MW's history, it
gives a good first-order approximation of what we may expect to observe.
At fixed apex direction, a larger travel velocity would correspond to a lower
inferred LG mass.
Figure~\ref{fig:mvsv} shows the effect of increasing the measured travel
velocity magnitude up to these predicted values and the impact on the
inferred mass of the LG for
each of the data sets used in this work (see Section~\ref{sec:data sets}).
The vertical lines in this figure show the LMC-induced travel velocities that
are predicted for three tracer distances in simulations from
~\cite{Garavito-Camargo2021b}.
Thus, future measurements of the travel velocity of the disk that use tracers at
larger distance around the MW stellar halo will likely lead to a lower inferred
LG mass.
We note again that the value of the eccentricity is derived assuming a matter-only universe (i.e., we neglect other cosmological effects in the orbit computation, as discussed above).

\subsection{Improved \mlg\ constraints from future observations}
\label{sec:discussion-futureobs}
The biggest source of uncertainty in our empirically inferred LG mass $\mlg$,
currently comes from the proper-motion measurements, which have signal-to-noise
ratios of just 3--4.
Future data releases from the \gaia\ Mission \citep{GaiaOverview2016} will lead
to more precise mean proper motions of M31 and thus more precise Timing Argument
constraints on the LG mass.
For example, between \gaia\ eDR3 and the end of the extended (10 yr) mission,
the expected individual-source proper-motion precision improvement for a $G=20$
source (i.e. an upper giant-branch star in M31) is a factor of $\sim$6.
Na\"ively scaling the proper-motion uncertainties of M31 as measured with \gaia\
\citep{Salomon2021} by a factor of 6 leads to a $\sim2\times$ improvement in the
$\mlg$ precision.
Of course, the true improvement of the mean M31 proper motion with improved
individual source kinematics could be even better than linear because more
sources will be detected and usable in the measurement.

\subsection{Reflex motion of M31}
\label{sec:discussion-m31}
While M31 has a massive satellite (M33) of comparable mass ratio to the
MW--LMC system, we do not expect there to be a significant reflex
velocity of M31's disk relative to its equivalent outer halo reference frame.
Recent work predicts that M33 is likely on first infall into the M31 halo and
has a much larger orbital pericenter than the MCs \citep[e.g.,][]{Patel2017a}.
Additionally, M31 has likely experienced other significant mergers, as
evidenced by the double nucleus and Giant Southern Stream
\citep[e.g.][]{Ibata:2001, Font:2006}, but these were likely lower mass-ratio
mergers \citep[e.g.][]{Gilbert:2019, Milo:2022} and thus will have less of an
impact on the bulk motion of the M31 disk.
Given current knowledge of the M31 system and uncertainties in the orbital
histories of its most massive satellites, here we neglect the reflex motion of
the M31 disk.
However, a measurement or upper limit on the M31 disk travel velocity would
enable further unbiased constraints on the LG mass.

\subsection{MW \& M31 Individual masses}
Reconciling techniques that compute the LG mass from the summed MW+M31 mass and
from the Timing Argument is not straightforward, but the two approaches
are complimentary.
Future measurements of the travel velocity at large Galactocentric distances
will likely exceed current measurements (see
Section~\ref{sec:discussion-impact}), which
directly implies a lower LG mass and may improve agreement between these two
general methods for estimating the LG mass.
Since constraints on the MW mass consistently find a mass of
~$\sim10^{12}\Msun$, if not slightly higher (see for reference
Table~\ref{table:masses}), our mass limits from the Timing Argument may begin to
place meaningful upper limits on the mass of the M31 system.

\subsection{The circularity of the MW--M31 orbit}
The Timing Argument is highly sensitive to the tangential motion of M31, and
larger proper motions will generically lead to lower eccentricities and higher
inferred LG masses.
Recent \gaia\ proper motions of M31 suggest the orbit of MW--M31 is less radial
than previously believed~\citep{vdm2019,Salomon2021}.
Neglecting the travel velocity ($\vtrav=0$ in above plots), we find that proper
motions from \textit{HST}~\citep{vdm2012} are consistent with a highly radial
orbit $e\sim0.96$, while proper motions from \textit{Gaia}~\citep{Salomon2021}
result in a slightly lower eccentricity of $e\sim0.88$.
As seen in Figure~~\ref{fig:mvsv}, we find that as the travel velocity
increases, the inferred eccentricity of the MW--M31 orbit decreases.
However, contrary to expectations, we find that the inferred LG mass
\textit{also} decreases.
This implies that the velocity contribution to the relative and transverse
velocity of M31 is dominant to the change in eccentricity.

Additionally, studies of the Bolshoi $N$-body cosmological simulation
by~\cite{Forero-Romero2013} find that typical LG analogs, when selected via mass and isolation criteria,
do not have completely radial orbits.
As with studies that measure a higher transverse velocity for M31, and thus a
slightly less-radial orbit~\citep{vdm2019,Salomon2021}, we find that the
decrease in the eccentricity due to the measured travel velocity makes the LG
less eccentric and, thus, more cosmologically typical.

\section{Summary and Conclusions}
\label{sec:summary}
Recent measurements of tracers in the outer MW stellar halo $(40<r<120\kpc)$
by~\cite{Petersen2021} measure an instantaneous differential ``travel velocity''
of the MW disk compared to the outer stellar halo.
The travel velocity has been inferred as primarily due to the response of the
MW halo to the recent infall of the LMC
~\cite[as shown in e.g.][]{Gomez2015,Garavito-Camargo:2019,Erkal2019,
Cunningham:2020,Petersen:2020,Garavito-Camargo2021b}.
In this work, we
study the effect of the travel velocity on the inferred LG mass from the
Timing Argument empirically for the first time.
This allows us to avoid modeling uncertainties in the LMC orbital
history and mass profile as well as uncertainties in the MW merger history.

We also consider three compilations of kinematic data for the distance and
proper motion of M31, and find a decrease in the inferred mass compared to
non-travel-velocity TA models. For each
data set, as follows:
\footnote{Errors correspond to the 68\% credible regions about
the median LG mass from the MCMC sampled posterior pdfs
of our model parameters.}
\begin{itemize}
  \item[-] For the ``\textit{vdMG08 Dist. + HST PM}'' data set -- the
  ~\cite{vdm2008} distance and \textit{HST} proper-motion
  measurements~\citep{Sohn:2012,vdm2012} -- we find a
  LG mass of $\mlg = 3.98^{+0.63}_{-0.47}\times10^{12}~\Msun$
  (including the measured travel velocity), and of
  $\mlg = 4.49^{+0.47}_{-0.42}\times10^{12}~\Msun$ when $\vtrav$=0.
  Thus, the inclusion of the that the travel velocity decreases the inferred
  LG mass by 11.36\%.
  \item[-] For the ``\textit{Cepheid Dist. + HST PM}'' data set -- an updated
  Cepheid distance measurement from~\cite{Li2021} and the same \textit{HST} PMs
  from above --
  we find a LG mass of $\mlg = 4.05^{+0.51}_{-0.34}\times10^{12}~\Msun$
  (including the measured travel velocity), and of
  $\mlg = 4.61^{+0.42}_{-0.22}\times10^{12}~\Msun$ when $\vtrav$=0.
  Thus, the inclusion of the that the travel velocity decreases the inferred
  LG mass by 12.01\%.

  \item[-] For the ``\textit{Cepheid Dist. + Gaia PM}'' data set -- a combination
  of the more recent Cepheid distance and latest \textit{Gaia} proper motion
  measurements of M31~\citep{Li2021,Salomon2021} --
  we find a LG mass of $\mlg = 4.54^{+0.77}_{-0.56}\times10^{12}~\Msun$
  (including the measured travel velocity), and of
  $\mlg = 5.09^{+0.72}_{-0.48}\times10^{12}~\Msun$ when $\vtrav$=0.
  Thus, the inclusion of the that the travel velocity decreases the inferred
  LG mass by 10.88\%.
\end{itemize}

Our conclusions can be summarized as follows:
\begin{itemize}
  \item \textbf{The measured travel velocity of the MW disk directly implies
  a reduced LG mass from the Timing Argument.}
  For the measured travel velocity of $32\pm4 \kms$ from~\cite{Petersen2021},
  the inferred LG mass is $\sim10-15\%$ lower than a system with a static MW
  halo ($\vtrav=0$).
  Using the same distance and proper motions as in the Timing Argument model
  of~\cite{vdm2012}, we find that the inclusion of the travel velocity yields
  a reduction in the LG mass of $\sim0.3\times 10^{12}\Msun$.
  \item \textbf{Higher travel velocity measurements will yield lower LG masses.}
  Simulations~\citep{Garavito-Camargo2021b} suggest that tracers at large
  Galactocentric distances $(60$--$100~\kpc)$ will yield larger measurements of
  the travel velocity.
  If the travel velocity is measured to be larger based on tracers at larger
  Galactocentric distances, this will result in a decrease in the inferred LG
  mass by an additional $5-12\%$.
  \item \textbf{The inferred eccentricity of the MW--M31 orbit is decreased by
  3--5\% when accounting for the measured travel velocity.}
  With a larger measured travel velocity, the inferred MW--M31 orbit would be
  less radial.
  The inferred eccentricity decreases by up to $\sim50\%$ for the largest travel
  velocities we consider ($\vtrav=100\kms$) compared to the static MW-halo model
  ($\vtrav=0$).
  Less radial orbits are cosmologically preferred, thus the travel velocity
  makes the LG more cosmologically typical.
  \item \textbf{Improvements in M31 proper motion measurements will improve
  Timing Argument mass precision.}
  With future data releases from the \gaia\ Mission, we expect the proper-motion
  uncertainties to improve by a factor of $\sim$2--3 for individual sources (and
  likely more for measurements of the mean proper motions of stellar systems
  and galaxies like M31).
  We artificially scaled the proper-motion errors in each data set and find an
  expected improvement in the uncertainty on the inferred LG mass by a factor of
  $2$--$2.5\times$ (see Section~\ref{sec:discussion-futureobs} for more detail).
\end{itemize}

This study highlights the importance of improved dynamical measurements within
the LV in the near future in order to accurately measure the dark
matter content of our LG.
It is critical to refine our measurements of the proper motion of M31 and to
measure the travel velocity of the MW disk with stellar tracers at
further Galactocentric distances.
These endeavors will (1) further refine estimates of the mass of the Local
Group, enabling studies to realistically place the LG in a cosmological
context, and (2) permit measurements of the travel velocity induced by the infall
of the LMC and other satellite galaxies relative to tracers at large
Galactocentric distances, which will establish a firm measurement of the LG mass via the Timing Argument, and thereby place meaningful limits on the
individual masses of the M31 and MW galaxy.

\begin{acknowledgements}

This project was started at the Big Apple Dynamics School (BADS) hosted by the
Flatiron Institute July--August 2021.
We greatly benefitted from discussions with the other students who attended the
BADS, and received helpful input from:
Kathryn Johnston (Columbia), Alex Riley (Texas A\&M), and Martin Weinberg
(University of Massachusetts at Amherst).
K.C. would like to thank Ekta Patel for sharing a collection of M31 mass
measurements from the literature.
This research made use of Astropy,\footnote{http://www.astropy.org} a
community-developed core Python package for Astronomy \citep{astropy:2013,
astropy:2018}.
K.C. and G.B. are supported by NSF CAREER AST-1941096 and NASA ATP 17-ATP17-0006.
M.S.P. acknowledges grant support from Segal ANR-19-CE31-0017 of the French Agence
Nationale de la Recherche (\url{https://secular-evolution.org}).

\end{acknowledgements}

\software{
    Arviz \citep{arviz},
    Astropy \citep{astropy:2013, astropy:2018},
    gala \citep{gala},
    IPython \citep{ipython},
    matplotlib \citep{matplotlib},
    numpy \citep{numpy},
    pymc3 \citep{Salvatier2016},
    scipy \citep{scipy}
}

\bibliography{refs}{}

\begin{thebibliography}{}
\expandafter\ifx\csname natexlab\endcsname\relax\def\natexlab#1{#1}\fi
\providecommand{\url}[1]{\href{#1}{#1}}
\providecommand{\dodoi}[1]{doi:~\href{http://doi.org/#1}{\nolinkurl{#1}}}
\providecommand{\doeprint}[1]{\href{http://ascl.net/#1}{\nolinkurl{http://ascl.net/#1}}}
\providecommand{\doarXiv}[1]{\href{https://arxiv.org/abs/#1}{\nolinkurl{https://arxiv.org/abs/#1}}}

\bibitem[{{Astropy Collaboration} {et~al.}(2013){Astropy Collaboration},
  {Robitaille}, {Tollerud}, {Greenfield}, {Droettboom}, {Bray}, {Aldcroft},
  {Davis}, {Ginsburg}, {Price-Whelan}, {Kerzendorf}, {Conley}, {Crighton},
  {Barbary}, {Muna}, {Ferguson}, {Grollier}, {Parikh}, {Nair}, {Unther},
  {Deil}, {Woillez}, {Conseil}, {Kramer}, {Turner}, {Singer}, {Fox}, {Weaver},
  {Zabalza}, {Edwards}, {Azalee Bostroem}, {Burke}, {Casey}, {Crawford},
  {Dencheva}, {Ely}, {Jenness}, {Labrie}, {Lim}, {Pierfederici}, {Pontzen},
  {Ptak}, {Refsdal}, {Servillat}, \& {Streicher}}]{astropy:2013}
{Astropy Collaboration}, {Robitaille}, T.~P., {Tollerud}, E.~J., {et~al.} 2013,
  \aap, 558, A33, \dodoi{10.1051/0004-6361/201322068}

\bibitem[{{Astropy Collaboration} {et~al.}(2018){Astropy Collaboration},
  {Price-Whelan}, {Sip{\H{o}}cz}, {G{\"u}nther}, {Lim}, {Crawford}, {Conseil},
  {Shupe}, {Craig}, {Dencheva}, {Ginsburg}, {Vand erPlas}, {Bradley},
  {P{\'e}rez-Su{\'a}rez}, {de Val-Borro}, {Aldcroft}, {Cruz}, {Robitaille},
  {Tollerud}, {Ardelean}, {Babej}, {Bach}, {Bachetti}, {Bakanov}, {Bamford},
  {Barentsen}, {Barmby}, {Baumbach}, {Berry}, {Biscani}, {Boquien}, {Bostroem},
  {Bouma}, {Brammer}, {Bray}, {Breytenbach}, {Buddelmeijer}, {Burke},
  {Calderone}, {Cano Rodr{\'\i}guez}, {Cara}, {Cardoso}, {Cheedella}, {Copin},
  {Corrales}, {Crichton}, {D'Avella}, {Deil}, {Depagne}, {Dietrich}, {Donath},
  {Droettboom}, {Earl}, {Erben}, {Fabbro}, {Ferreira}, {Finethy}, {Fox},
  {Garrison}, {Gibbons}, {Goldstein}, {Gommers}, {Greco}, {Greenfield},
  {Groener}, {Grollier}, {Hagen}, {Hirst}, {Homeier}, {Horton}, {Hosseinzadeh},
  {Hu}, {Hunkeler}, {Ivezi{\'c}}, {Jain}, {Jenness}, {Kanarek}, {Kendrew},
  {Kern}, {Kerzendorf}, {Khvalko}, {King}, {Kirkby}, {Kulkarni}, {Kumar},
  {Lee}, {Lenz}, {Littlefair}, {Ma}, {Macleod}, {Mastropietro}, {McCully},
  {Montagnac}, {Morris}, {Mueller}, {Mumford}, {Muna}, {Murphy}, {Nelson},
  {Nguyen}, {Ninan}, {N{\"o}the}, {Ogaz}, {Oh}, {Parejko}, {Parley}, {Pascual},
  {Patil}, {Patil}, {Plunkett}, {Prochaska}, {Rastogi}, {Reddy Janga},
  {Sabater}, {Sakurikar}, {Seifert}, {Sherbert}, {Sherwood-Taylor}, {Shih},
  {Sick}, {Silbiger}, {Singanamalla}, {Singer}, {Sladen}, {Sooley},
  {Sornarajah}, {Streicher}, {Teuben}, {Thomas}, {Tremblay}, {Turner},
  {Terr{\'o}n}, {van Kerkwijk}, {de la Vega}, {Watkins}, {Weaver}, {Whitmore},
  {Woillez}, {Zabalza}, \& {Astropy Contributors}}]{astropy:2018}
{Astropy Collaboration}, {Price-Whelan}, A.~M., {Sip{\H{o}}cz}, B.~M., {et~al.}
  2018, \aj, 156, 123, \dodoi{10.3847/1538-3881/aabc4f}

\bibitem[{{Benisty} {et~al.}(2022){Benisty}, {Vasiliev}, {Evans}, {Davis},
  {Hartl}, \& {Strigari}}]{Benisty2022}
{Benisty}, D., {Vasiliev}, E., {Evans}, N.~W., {et~al.} 2022, arXiv e-prints,
  arXiv:2202.00033.
\newblock \doarXiv{2202.00033}

\bibitem[{{Bennett} \& {Bovy}(2019)}]{Bennett2019}
{Bennett}, M., \& {Bovy}, J. 2019, \mnras, 482, 1417,
  \dodoi{10.1093/mnras/sty2813}

\bibitem[{{Besla} {et~al.}(2018){Besla}, {Patton}, {Stierwalt},
  {Rodriguez-Gomez}, {Patel}, {Kallivayalil}, {Johnson}, {Pearson}, {Privon},
  \& {Putman}}]{Besla2018}
{Besla}, G., {Patton}, D.~R., {Stierwalt}, S., {et~al.} 2018, \mnras, 480,
  3376, \dodoi{10.1093/mnras/sty2041}

\bibitem[{{Carlesi} {et~al.}(2017){Carlesi}, {Hoffman}, {Sorce}, \&
  {Gottl{\"o}ber}}]{Carlesi2017}
{Carlesi}, E., {Hoffman}, Y., {Sorce}, J.~G., \& {Gottl{\"o}ber}, S. 2017,
  \mnras, 465, 4886, \dodoi{10.1093/mnras/stw3073}

\bibitem[{{Conroy} {et~al.}(2021){Conroy}, {Naidu}, {Garavito-Camargo},
  {Besla}, {Zaritsky}, {Bonaca}, \& {Johnson}}]{Conroy:2021}
{Conroy}, C., {Naidu}, R.~P., {Garavito-Camargo}, N., {et~al.} 2021, \nat, 592,
  534, \dodoi{10.1038/s41586-021-03385-7}

\bibitem[{{Courteau} \& {van den Bergh}(1999)}]{Courteau1999}
{Courteau}, S., \& {van den Bergh}, S. 1999, \aj, 118, 337,
  \dodoi{10.1086/300942}

\bibitem[{{Cunningham} {et~al.}(2020){Cunningham}, {Garavito-Camargo},
  {Deason}, {Johnston}, {Erkal}, {Laporte}, {Besla}, {Luger}, \&
  {Sanderson}}]{Cunningham:2020}
{Cunningham}, E.~C., {Garavito-Camargo}, N., {Deason}, A.~J., {et~al.} 2020,
  \apj, 898, 4, \dodoi{10.3847/1538-4357/ab9b88}

\bibitem[{{Deason} {et~al.}(2021){Deason}, {Erkal}, {Belokurov}, {Fattahi},
  {G{\'o}mez}, {Grand}, {Pakmor}, {Xue}, {Liu}, {Yang}, {Zhang}, \&
  {Zhao}}]{Deason:2021}
{Deason}, A.~J., {Erkal}, D., {Belokurov}, V., {et~al.} 2021, \mnras, 501,
  5964, \dodoi{10.1093/mnras/staa3984}

\bibitem[{{Diaz} {et~al.}(2014){Diaz}, {Koposov}, {Irwin}, {Belokurov}, \&
  {Evans}}]{Diaz2014}
{Diaz}, J.~D., {Koposov}, S.~E., {Irwin}, M., {Belokurov}, V., \& {Evans},
  N.~W. 2014, \mnras, 443, 1688, \dodoi{10.1093/mnras/stu1210}

\bibitem[{{Dooley} {et~al.}(2017){Dooley}, {Peter}, {Yang}, {Willman},
  {Griffen}, \& {Frebel}}]{Dooley2017}
{Dooley}, G.~A., {Peter}, A. H.~G., {Yang}, T., {et~al.} 2017, \mnras, 471,
  4894, \dodoi{10.1093/mnras/stx1900}

\bibitem[{{Drimmel} \& {Poggio}(2018)}]{Drimmel2018}
{Drimmel}, R., \& {Poggio}, E. 2018, Research Notes of the American
  Astronomical Society, 2, 210, \dodoi{10.3847/2515-5172/aaef8b}

\bibitem[{{Eadie} \& {Juri{\'c}}(2019)}]{Eadie:2019}
{Eadie}, G., \& {Juri{\'c}}, M. 2019, \apj, 875, 159,
  \dodoi{10.3847/1538-4357/ab0f97}

\bibitem[{{Erkal} {et~al.}(2020){Erkal}, {Belokurov}, \& {Parkin}}]{Erkal2020}
{Erkal}, D., {Belokurov}, V.~A., \& {Parkin}, D.~L. 2020, \mnras, 498, 5574,
  \dodoi{10.1093/mnras/staa2840}

\bibitem[{{Erkal} {et~al.}(2019){Erkal}, {Belokurov}, {Laporte}, {Koposov},
  {Li}, {Grillmair}, {Kallivayalil}, {Price-Whelan}, {Evans}, {Hawkins},
  {Hendel}, {Mateu}, {Navarro}, {del Pino}, {Slater}, {Sohn}, \& {Orphan Aspen
  Treasury Collaboration}}]{Erkal2019}
{Erkal}, D., {Belokurov}, V., {Laporte}, C.~F.~P., {et~al.} 2019, \mnras, 487,
  2685, \dodoi{10.1093/mnras/stz1371}

\bibitem[{{Erkal} {et~al.}(2021){Erkal}, {Deason}, {Belokurov}, {Xue},
  {Koposov}, {Bird}, {Liu}, {Simion}, {Yang}, {Zhang}, \& {Zhao}}]{Erkal:2021}
{Erkal}, D., {Deason}, A.~J., {Belokurov}, V., {et~al.} 2021, \mnras, 506,
  2677, \dodoi{10.1093/mnras/stab1828}

\bibitem[{{Fardal} {et~al.}(2013){Fardal}, {Weinberg}, {Babul}, {Irwin},
  {Guhathakurta}, {Gilbert}, {Ferguson}, {Ibata}, {Lewis}, {Tanvir}, \&
  {Huxor}}]{Fardal2013}
{Fardal}, M.~A., {Weinberg}, M.~D., {Babul}, A., {et~al.} 2013, \mnras, 434,
  2779, \dodoi{10.1093/mnras/stt1121}

\bibitem[{{Fillingham} {et~al.}(2018){Fillingham}, {Cooper}, {Boylan-Kolchin},
  {Bullock}, {Garrison-Kimmel}, \& {Wheeler}}]{Fillingham:2018}
{Fillingham}, S.~P., {Cooper}, M.~C., {Boylan-Kolchin}, M., {et~al.} 2018,
  \mnras, 477, 4491, \dodoi{10.1093/mnras/sty958}

\bibitem[{{Font} {et~al.}(2006){Font}, {Johnston}, {Guhathakurta}, {Majewski},
  \& {Rich}}]{Font:2006}
{Font}, A.~S., {Johnston}, K.~V., {Guhathakurta}, P., {Majewski}, S.~R., \&
  {Rich}, R.~M. 2006, \aj, 131, 1436, \dodoi{10.1086/499564}

\bibitem[{{Forero-Romero} {et~al.}(2013){Forero-Romero}, {Hoffman},
  {Bustamante}, {Gottl{\"o}ber}, \& {Yepes}}]{Forero-Romero2013}
{Forero-Romero}, J.~E., {Hoffman}, Y., {Bustamante}, S., {Gottl{\"o}ber}, S.,
  \& {Yepes}, G. 2013, \apjl, 767, L5, \dodoi{10.1088/2041-8205/767/1/L5}

\bibitem[{{Fritz} {et~al.}(2020){Fritz}, {Di Cintio}, {Battaglia}, {Brook}, \&
  {Taibi}}]{Fritz:2020}
{Fritz}, T.~K., {Di Cintio}, A., {Battaglia}, G., {Brook}, C., \& {Taibi}, S.
  2020, \mnras, 494, 5178, \dodoi{10.1093/mnras/staa1040}

\bibitem[{{Gaia Collaboration} {et~al.}(2016){Gaia Collaboration}, {Prusti},
  {de Bruijne}, {Brown}, {Vallenari}, {Babusiaux}, {Bailer-Jones}, {Bastian},
  {Biermann}, {Evans}, {Eyer}, {Jansen}, {Jordi}, {Klioner}, {Lammers},
  {Lindegren}, {Luri}, {Mignard}, {Milligan}, {Panem}, {Poinsignon},
  {Pourbaix}, {Randich}, {Sarri}, {Sartoretti}, {Siddiqui}, {Soubiran},
  {Valette}, {van Leeuwen}, {Walton}, {Aerts}, {Arenou}, {Cropper}, {Drimmel},
  {H{\o}g}, {Katz}, {Lattanzi}, {O'Mullane}, {Grebel}, {Holland}, {Huc},
  {Passot}, {Bramante}, {Cacciari}, {Casta{\~n}eda}, {Chaoul}, {Cheek}, {De
  Angeli}, {Fabricius}, {Guerra}, {Hern{\'a}ndez}, {Jean-Antoine-Piccolo},
  {Masana}, {Messineo}, {Mowlavi}, {Nienartowicz}, {Ord{\'o}{\~n}ez-Blanco},
  {Panuzzo}, {Portell}, {Richards}, {Riello}, {Seabroke}, {Tanga},
  {Th{\'e}venin}, {Torra}, {Els}, {Gracia-Abril}, {Comoretto},
  {Garcia-Reinaldos}, {Lock}, {Mercier}, {Altmann}, {Andrae}, {Astraatmadja},
  {Bellas-Velidis}, {Benson}, {Berthier}, {Blomme}, {Busso}, {Carry},
  {Cellino}, {Clementini}, {Cowell}, {Creevey}, {Cuypers}, {Davidson}, {De
  Ridder}, {de Torres}, {Delchambre}, {Dell'Oro}, {Ducourant}, {Fr{\'e}mat},
  {Garc{\'\i}a-Torres}, {Gosset}, {Halbwachs}, {Hambly}, {Harrison}, {Hauser},
  {Hestroffer}, {Hodgkin}, {Huckle}, {Hutton}, {Jasniewicz}, {Jordan},
  {Kontizas}, {Korn}, {Lanzafame}, {Manteiga}, {Moitinho}, {Muinonen},
  {Osinde}, {Pancino}, {Pauwels}, {Petit}, {Recio-Blanco}, {Robin}, {Sarro},
  {Siopis}, {Smith}, {Smith}, {Sozzetti}, {Thuillot}, {van Reeven}, {Viala},
  {Abbas}, {Abreu Aramburu}, {Accart}, {Aguado}, {Allan}, {Allasia},
  {Altavilla}, {{\'A}lvarez}, {Alves}, {Anderson}, {Andrei}, {Anglada Varela},
  {Antiche}, {Antoja}, {Ant{\'o}n}, {Arcay}, {Atzei}, {Ayache}, {Bach},
  {Baker}, {Balaguer-N{\'u}{\~n}ez}, {Barache}, {Barata}, {Barbier}, {Barblan},
  {Baroni}, {Barrado y Navascu{\'e}s}, {Barros}, {Barstow}, {Becciani},
  {Bellazzini}, {Bellei}, {Bello Garc{\'\i}a}, {Belokurov}, {Bendjoya},
  {Berihuete}, {Bianchi}, {Bienaym{\'e}}, {Billebaud}, {Blagorodnova},
  {Blanco-Cuaresma}, {Boch}, {Bombrun}, {Borrachero}, {Bouquillon}, {Bourda},
  {Bouy}, {Bragaglia}, {Breddels}, {Brouillet}, {Br{\"u}semeister},
  {Bucciarelli}, {Budnik}, {Burgess}, {Burgon}, {Burlacu}, {Busonero}, {Buzzi},
  {Caffau}, {Cambras}, {Campbell}, {Cancelliere}, {Cantat-Gaudin}, {Carlucci},
  {Carrasco}, {Castellani}, {Charlot}, {Charnas}, {Charvet}, {Chassat},
  {Chiavassa}, {Clotet}, {Cocozza}, {Collins}, {Collins}, {Costigan}, {Crifo},
  {Cross}, {Crosta}, {Crowley}, {Dafonte}, {Damerdji}, {Dapergolas}, {David},
  {David}, {De Cat}, {de Felice}, {de Laverny}, {De Luise}, {De March}, {de
  Martino}, {de Souza}, {Debosscher}, {del Pozo}, {Delbo}, {Delgado},
  {Delgado}, {di Marco}, {Di Matteo}, {Diakite}, {Distefano}, {Dolding}, {Dos
  Anjos}, {Drazinos}, {Dur{\'a}n}, {Dzigan}, {Ecale}, {Edvardsson}, {Enke},
  {Erdmann}, {Escolar}, {Espina}, {Evans}, {Eynard Bontemps}, {Fabre},
  {Fabrizio}, {Faigler}, {Falc{\~a}o}, {Farr{\`a}s Casas}, {Faye}, {Federici},
  {Fedorets}, {Fern{\'a}ndez-Hern{\'a}ndez}, {Fernique}, {Fienga}, {Figueras},
  {Filippi}, {Findeisen}, {Fonti}, {Fouesneau}, {Fraile}, {Fraser}, {Fuchs},
  {Furnell}, {Gai}, {Galleti}, {Galluccio}, {Garabato}, {Garc{\'\i}a-Sedano},
  {Gar{\'e}}, {Garofalo}, {Garralda}, {Gavras}, {Gerssen}, {Geyer}, {Gilmore},
  {Girona}, {Giuffrida}, {Gomes}, {Gonz{\'a}lez-Marcos},
  {Gonz{\'a}lez-N{\'u}{\~n}ez}, {Gonz{\'a}lez-Vidal}, {Granvik}, {Guerrier},
  {Guillout}, {Guiraud}, {G{\'u}rpide}, {Guti{\'e}rrez-S{\'a}nchez}, {Guy},
  {Haigron}, {Hatzidimitriou}, {Haywood}, {Heiter}, {Helmi}, {Hobbs},
  {Hofmann}, {Holl}, {Holland}, {Hunt}, {Hypki}, {Icardi}, {Irwin}, {Jevardat
  de Fombelle}, {Jofr{\'e}}, {Jonker}, {Jorissen}, {Julbe}, {Karampelas},
  {Kochoska}, {Kohley}, {Kolenberg}, {Kontizas}, {Koposov}, {Kordopatis},
  {Koubsky}, {Kowalczyk}, {Krone-Martins}, {Kudryashova}, {Kull}, {Bachchan},
  {Lacoste-Seris}, {Lanza}, {Lavigne}, {Le Poncin-Lafitte}, {Lebreton},
  {Lebzelter}, {Leccia}, {Leclerc}, {Lecoeur-Taibi}, {Lemaitre}, {Lenhardt},
  {Leroux}, {Liao}, {Licata}, {Lindstr{\o}m}, {Lister}, {Livanou}, {Lobel},
  {L{\"o}ffler}, {L{\'o}pez}, {Lopez-Lozano}, {Lorenz}, {Loureiro},
  {MacDonald}, {Magalh{\~a}es Fernandes}, {Managau}, {Mann}, {Mantelet},
  {Marchal}, {Marchant}, {Marconi}, {Marie}, {Marinoni}, {Marrese},
  {Marschalk{\'o}}, {Marshall}, {Mart{\'\i}n-Fleitas}, {Martino}, {Mary},
  {Matijevi{\v{c}}}, {Mazeh}, {McMillan}, {Messina}, {Mestre}, {Michalik},
  {Millar}, {Miranda}, {Molina}, {Molinaro}, {Molinaro}, {Moln{\'a}r},
  {Moniez}, {Montegriffo}, {Monteiro}, {Mor}, {Mora}, {Morbidelli}, {Morel},
  {Morgenthaler}, {Morley}, {Morris}, {Mulone}, {Muraveva}, {Musella},
  {Narbonne}, {Nelemans}, {Nicastro}, {Noval}, {Ord{\'e}novic},
  {Ordieres-Mer{\'e}}, {Osborne}, {Pagani}, {Pagano}, {Pailler}, {Palacin},
  {Palaversa}, {Parsons}, {Paulsen}, {Pecoraro}, {Pedrosa}, {Pentik{\"a}inen},
  {Pereira}, {Pichon}, {Piersimoni}, {Pineau}, {Plachy}, {Plum}, {Poujoulet},
  {Pr{\v{s}}a}, {Pulone}, {Ragaini}, {Rago}, {Rambaux}, {Ramos-Lerate},
  {Ranalli}, {Rauw}, {Read}, {Regibo}, {Renk}, {Reyl{\'e}}, {Ribeiro},
  {Rimoldini}, {Ripepi}, {Riva}, {Rixon}, {Roelens}, {Romero-G{\'o}mez},
  {Rowell}, {Royer}, {Rudolph}, {Ruiz-Dern}, {Sadowski}, {Sagrist{\`a}
  Sell{\'e}s}, {Sahlmann}, {Salgado}, {Salguero}, {Sarasso}, {Savietto},
  {Schnorhk}, {Schultheis}, {Sciacca}, {Segol}, {Segovia}, {Segransan},
  {Serpell}, {Shih}, {Smareglia}, {Smart}, {Smith}, {Solano}, {Solitro},
  {Sordo}, {Soria Nieto}, {Souchay}, {Spagna}, {Spoto}, {Stampa}, {Steele},
  {Steidelm{\"u}ller}, {Stephenson}, {Stoev}, {Suess}, {S{\"u}veges}, {Surdej},
  {Szabados}, {Szegedi-Elek}, {Tapiador}, {Taris}, {Tauran}, {Taylor},
  {Teixeira}, {Terrett}, {Tingley}, {Trager}, {Turon}, {Ulla}, {Utrilla},
  {Valentini}, {van Elteren}, {Van Hemelryck}, {van Leeuwen}, {Varadi},
  {Vecchiato}, {Veljanoski}, {Via}, {Vicente}, {Vogt}, {Voss}, {Votruba},
  {Voutsinas}, {Walmsley}, {Weiler}, {Weingrill}, {Werner}, {Wevers},
  {Whitehead}, {Wyrzykowski}, {Yoldas}, {{\v{Z}}erjal}, {Zucker}, {Zurbach},
  {Zwitter}, {Alecu}, {Allen}, {Allende Prieto}, {Amorim},
  {Anglada-Escud{\'e}}, {Arsenijevic}, {Azaz}, {Balm}, {Beck}, {Bernstein},
  {Bigot}, {Bijaoui}, {Blasco}, {Bonfigli}, {Bono}, {Boudreault}, {Bressan},
  {Brown}, {Brunet}, {Bunclark}, {Buonanno}, {Butkevich}, {Carret}, {Carrion},
  {Chemin}, {Ch{\'e}reau}, {Corcione}, {Darmigny}, {de Boer}, {de Teodoro}, {de
  Zeeuw}, {Delle Luche}, {Domingues}, {Dubath}, {Fodor}, {Fr{\'e}zouls},
  {Fries}, {Fustes}, {Fyfe}, {Gallardo}, {Gallegos}, {Gardiol}, {Gebran},
  {Gomboc}, {G{\'o}mez}, {Grux}, {Gueguen}, {Heyrovsky}, {Hoar}, {Iannicola},
  {Isasi Parache}, {Janotto}, {Joliet}, {Jonckheere}, {Keil}, {Kim},
  {Klagyivik}, {Klar}, {Knude}, {Kochukhov}, {Kolka}, {Kos}, {Kutka}, {Lainey},
  {LeBouquin}, {Liu}, {Loreggia}, {Makarov}, {Marseille}, {Martayan},
  {Martinez-Rubi}, {Massart}, {Meynadier}, {Mignot}, {Munari}, {Nguyen},
  {Nordlander}, {Ocvirk}, {O'Flaherty}, {Olias Sanz}, {Ortiz}, {Osorio},
  {Oszkiewicz}, {Ouzounis}, {Palmer}, {Park}, {Pasquato}, {Peltzer}, {Peralta},
  {P{\'e}turaud}, {Pieniluoma}, {Pigozzi}, {Poels}, {Prat}, {Prod'homme},
  {Raison}, {Rebordao}, {Risquez}, {Rocca-Volmerange}, {Rosen}, {Ruiz-Fuertes},
  {Russo}, {Sembay}, {Serraller Vizcaino}, {Short}, {Siebert}, {Silva},
  {Sinachopoulos}, {Slezak}, {Soffel}, {Sosnowska}, {Strai{\v{z}}ys}, {ter
  Linden}, {Terrell}, {Theil}, {Tiede}, {Troisi}, {Tsalmantza}, {Tur},
  {Vaccari}, {Vachier}, {Valles}, {Van Hamme}, {Veltz}, {Virtanen}, {Wallut},
  {Wichmann}, {Wilkinson}, {Ziaeepour}, \& {Zschocke}}]{GaiaOverview2016}
{Gaia Collaboration}, {Prusti}, T., {de Bruijne}, J.~H.~J., {et~al.} 2016,
  \aap, 595, A1, \dodoi{10.1051/0004-6361/201629272}

\bibitem[{{Garavito-Camargo} {et~al.}(2019){Garavito-Camargo}, {Besla},
  {Laporte}, {Johnston}, {G{\'o}mez}, \& {Watkins}}]{Garavito-Camargo:2019}
{Garavito-Camargo}, N., {Besla}, G., {Laporte}, C. F.~P., {et~al.} 2019, \apj,
  884, 51, \dodoi{10.3847/1538-4357/ab32eb}

\bibitem[{{Garavito-Camargo} {et~al.}(2021){Garavito-Camargo}, {Besla},
  {Laporte}, {Price-Whelan}, {Cunningham}, {Johnston}, {Weinberg}, \&
  {G{\'o}mez}}]{Garavito-Camargo2021b}
---. 2021, \apj, 919, 109, \dodoi{10.3847/1538-4357/ac0b44}

\bibitem[{{Garrison-Kimmel} {et~al.}(2019{\natexlab{a}}){Garrison-Kimmel},
  {Hopkins}, {Wetzel}, {Bullock}, {Boylan-Kolchin}, {Kere{\v{s}}},
  {Faucher-Gigu{\`e}re}, {El-Badry}, {Lamberts}, {Quataert}, \&
  {Sanderson}}]{Garrison-Kimmel:2019a}
{Garrison-Kimmel}, S., {Hopkins}, P.~F., {Wetzel}, A., {et~al.}
  2019{\natexlab{a}}, \mnras, 487, 1380, \dodoi{10.1093/mnras/stz1317}

\bibitem[{{Garrison-Kimmel} {et~al.}(2019{\natexlab{b}}){Garrison-Kimmel},
  {Wetzel}, {Hopkins}, {Sanderson}, {El-Badry}, {Graus}, {Chan}, {Feldmann},
  {Boylan-Kolchin}, {Hayward}, {Bullock}, {Fitts}, {Samuel}, {Wheeler},
  {Kere{\v{s}}}, \& {Faucher-Gigu{\`e}re}}]{Garrison-Kimmel:2019b}
{Garrison-Kimmel}, S., {Wetzel}, A., {Hopkins}, P.~F., {et~al.}
  2019{\natexlab{b}}, \mnras, 489, 4574, \dodoi{10.1093/mnras/stz2507}

\bibitem[{{Gelman} \& {Rubin}(1992)}]{GelmanRubin1992}
{Gelman}, A., \& {Rubin}, D.~B. 1992, Statistical Science, 7, 457,
  \dodoi{10.1214/ss/1177011136}

\bibitem[{{Gilbert} {et~al.}(2019){Gilbert}, {Kirby}, {Escala}, {Wojno},
  {Kalirai}, \& {Guhathakurta}}]{Gilbert:2019}
{Gilbert}, K.~M., {Kirby}, E.~N., {Escala}, I., {et~al.} 2019, \apj, 883, 128,
  \dodoi{10.3847/1538-4357/ab3807}

\bibitem[{{G{\'o}mez} {et~al.}(2015){G{\'o}mez}, {Besla}, {Carpintero},
  {Villalobos}, {O'Shea}, \& {Bell}}]{Gomez2015}
{G{\'o}mez}, F.~A., {Besla}, G., {Carpintero}, D.~D., {et~al.} 2015, \apj, 802,
  128, \dodoi{10.1088/0004-637X/802/2/128}

\bibitem[{{Gonz{\'a}lez} {et~al.}(2014){Gonz{\'a}lez}, {Kravtsov}, \&
  {Gnedin}}]{Gonzalez2014}
{Gonz{\'a}lez}, R.~E., {Kravtsov}, A.~V., \& {Gnedin}, N.~Y. 2014, \apj, 793,
  91, \dodoi{10.1088/0004-637X/793/2/91}

\bibitem[{{Gravity Collaboration} {et~al.}(2018){Gravity Collaboration},
  {Abuter}, {Amorim}, {Anugu}, {Baub{\"o}ck}, {Benisty}, {Berger}, {Blind},
  {Bonnet}, {Brandner}, {Buron}, {Collin}, {Chapron}, {Cl{\'e}net}, {Coud{\'e}
  Du Foresto}, {de Zeeuw}, {Deen}, {Delplancke-Str{\"o}bele}, {Dembet},
  {Dexter}, {Duvert}, {Eckart}, {Eisenhauer}, {Finger}, {F{\"o}rster
  Schreiber}, {F{\'e}dou}, {Garcia}, {Garcia Lopez}, {Gao}, {Gendron},
  {Genzel}, {Gillessen}, {Gordo}, {Habibi}, {Haubois}, {Haug}, {Hau{\ss}mann},
  {Henning}, {Hippler}, {Horrobin}, {Hubert}, {Hubin}, {Jimenez Rosales},
  {Jochum}, {Jocou}, {Kaufer}, {Kellner}, {Kendrew}, {Kervella}, {Kok},
  {Kulas}, {Lacour}, {Lapeyr{\`e}re}, {Lazareff}, {Le Bouquin}, {L{\'e}na},
  {Lippa}, {Lenzen}, {M{\'e}rand}, {M{\"u}ler}, {Neumann}, {Ott}, {Palanca},
  {Paumard}, {Pasquini}, {Perraut}, {Perrin}, {Pfuhl}, {Plewa}, {Rabien},
  {Ram{\'\i}rez}, {Ramos}, {Rau}, {Rodr{\'\i}guez-Coira}, {Rohloff}, {Rousset},
  {Sanchez-Bermudez}, {Scheithauer}, {Sch{\"o}ller}, {Schuler}, {Spyromilio},
  {Straub}, {Straubmeier}, {Sturm}, {Tacconi}, {Tristram}, {Vincent}, {von
  Fellenberg}, {Wank}, {Waisberg}, {Widmann}, {Wieprecht}, {Wiest},
  {Wiezorrek}, {Woillez}, {Yazici}, {Ziegler}, \& {Zins}}]{GravityCollab2018}
{Gravity Collaboration}, {Abuter}, R., {Amorim}, A., {et~al.} 2018, \aap, 615,
  L15, \dodoi{10.1051/0004-6361/201833718}

\bibitem[{Harris {et~al.}(2020)Harris, Millman, van~der Walt, Gommers,
  Virtanen, Cournapeau, Wieser, Taylor, Berg, Smith, Kern, Picus, Hoyer, van
  Kerkwijk, Brett, Haldane, del R{\'{i}}o, Wiebe, Peterson,
  G{\'{e}}rard-Marchant, Sheppard, Reddy, Weckesser, Abbasi, Gohlke, \&
  Oliphant}]{numpy}
Harris, C.~R., Millman, K.~J., van~der Walt, S.~J., {et~al.} 2020, Nature, 585,
  357, \dodoi{10.1038/s41586-020-2649-2}

\bibitem[{{Hartl} \& {Strigari}(2022)}]{Hartl2022}
{Hartl}, O.~V., \& {Strigari}, L.~E. 2022, \mnras, 511, 6193,
  \dodoi{10.1093/mnras/stac413}

\bibitem[{Homan \& Gelman(2014)}]{Homan2014}
Homan, M.~D., \& Gelman, A. 2014, J. Mach. Learn. Res., 15, 1593–1623

\bibitem[{Hunter(2007)}]{matplotlib}
Hunter, J.~D. 2007, Computing in Science \& Engineering, 9, 90,
  \dodoi{10.1109/MCSE.2007.55}

\bibitem[{{Ibata} {et~al.}(2001){Ibata}, {Irwin}, {Lewis}, {Ferguson}, \&
  {Tanvir}}]{Ibata:2001}
{Ibata}, R., {Irwin}, M., {Lewis}, G., {Ferguson}, A. M.~N., \& {Tanvir}, N.
  2001, \nat, 412, 49.
\newblock \doarXiv{astro-ph/0107090}

\bibitem[{{Jarosik} {et~al.}(2011){Jarosik}, {Bennett}, {Dunkley}, {Gold},
  {Greason}, {Halpern}, {Hill}, {Hinshaw}, {Kogut}, {Komatsu}, {Larson},
  {Limon}, {Meyer}, {Nolta}, {Odegard}, {Page}, {Smith}, {Spergel}, {Tucker},
  {Weiland}, {Wollack}, \& {Wright}}]{Jarosik2011}
{Jarosik}, N., {Bennett}, C.~L., {Dunkley}, J., {et~al.} 2011, \apjs, 192, 14,
  \dodoi{10.1088/0067-0049/192/2/14}

\bibitem[{{Kahn} \& {Woltjer}(1959)}]{Kahn1959}
{Kahn}, F.~D., \& {Woltjer}, L. 1959, \apj, 130, 705, \dodoi{10.1086/146762}

\bibitem[{{Kroeker} \& {Carlberg}(1991)}]{Kroeker1991}
{Kroeker}, T.~L., \& {Carlberg}, R.~G. 1991, \apj, 376, 1,
  \dodoi{10.1086/170249}

\bibitem[{Kumar {et~al.}(2019)Kumar, Carroll, Hartikainen, \& Martin}]{arviz}
Kumar, R., Carroll, C., Hartikainen, A., \& Martin, O. 2019, Journal of Open
  Source Software, 4, 1143, \dodoi{10.21105/joss.01143}

\bibitem[{{Laporte} {et~al.}(2018{\natexlab{a}}){Laporte}, {G{\'o}mez},
  {Besla}, {Johnston}, \& {Garavito-Camargo}}]{Laporte:2018a}
{Laporte}, C. F.~P., {G{\'o}mez}, F.~A., {Besla}, G., {Johnston}, K.~V., \&
  {Garavito-Camargo}, N. 2018{\natexlab{a}}, \mnras, 473, 1218,
  \dodoi{10.1093/mnras/stx2146}

\bibitem[{{Laporte} {et~al.}(2018{\natexlab{b}}){Laporte}, {Johnston},
  {G{\'o}mez}, {Garavito-Camargo}, \& {Besla}}]{Laporte:2018b}
{Laporte}, C. F.~P., {Johnston}, K.~V., {G{\'o}mez}, F.~A., {Garavito-Camargo},
  N., \& {Besla}, G. 2018{\natexlab{b}}, \mnras, 481, 286,
  \dodoi{10.1093/mnras/sty1574}

\bibitem[{Lemos {et~al.}(2021)Lemos, Jeffrey, Whiteway, Lahav, Libeskind, \&
  Hoffman}]{Lemos2021}
Lemos, P., Jeffrey, N., Whiteway, L., {et~al.} 2021, Phys. Rev. D, 103, 023009,
  \dodoi{10.1103/PhysRevD.103.023009}

\bibitem[{{Li} {et~al.}(2021){Li}, {Riess}, {Busch}, {Casertano}, {Macri}, \&
  {Yuan}}]{Li2021}
{Li}, S., {Riess}, A.~G., {Busch}, M.~P., {et~al.} 2021, \apj, 920, 84,
  \dodoi{10.3847/1538-4357/ac1597}

\bibitem[{{Li} \& {White}(2008)}]{LiWhite2008}
{Li}, Y.-S., \& {White}, S. D.~M. 2008, \mnras, 384, 1459,
  \dodoi{10.1111/j.1365-2966.2007.12748.x}

\bibitem[{{Lynden-Bell}(1981)}]{Lynden-Bell:1981}
{Lynden-Bell}, D. 1981, The Observatory, 101, 111

\bibitem[{{Marinacci} {et~al.}(2017){Marinacci}, {Grand}, {Pakmor}, {Springel},
  {G{\'o}mez}, {Frenk}, \& {White}}]{Marinacci:2017}
{Marinacci}, F., {Grand}, R. J.~J., {Pakmor}, R., {et~al.} 2017, \mnras, 466,
  3859, \dodoi{10.1093/mnras/stw3366}

\bibitem[{{McLeod} {et~al.}(2017){McLeod}, {Libeskind}, {Lahav}, \&
  {Hoffman}}]{McLeod2017}
{McLeod}, M., {Libeskind}, N., {Lahav}, O., \& {Hoffman}, Y. 2017, \jcap, 2017,
  034, \dodoi{10.1088/1475-7516/2017/12/034}

\bibitem[{{McMillan}(2011)}]{McMillan2011}
{McMillan}, P.~J. 2011, \mnras, 414, 2446,
  \dodoi{10.1111/j.1365-2966.2011.18564.x}

\bibitem[{{Milo{\v{s}}evi{\'c}} {et~al.}(2022){Milo{\v{s}}evi{\'c}},
  {Mi{\'c}i{\'c}}, \& {Lewis}}]{Milo:2022}
{Milo{\v{s}}evi{\'c}}, S., {Mi{\'c}i{\'c}}, M., \& {Lewis}, G.~F. 2022, \mnras,
  \dodoi{10.1093/mnras/stac249}

\bibitem[{{Partridge} {et~al.}(2013){Partridge}, {Lahav}, \&
  {Hoffman}}]{Partridge:2013}
{Partridge}, C., {Lahav}, O., \& {Hoffman}, Y. 2013, \mnras, 436, L45,
  \dodoi{10.1093/mnrasl/slt109}

\bibitem[{{Patel} {et~al.}(2017{\natexlab{a}}){Patel}, {Besla}, \&
  {Mandel}}]{Patel2017b}
{Patel}, E., {Besla}, G., \& {Mandel}, K. 2017{\natexlab{a}}, \mnras, 468,
  3428, \dodoi{10.1093/mnras/stx698}

\bibitem[{{Patel} {et~al.}(2018){Patel}, {Besla}, {Mandel}, \&
  {Sohn}}]{Patel2018}
{Patel}, E., {Besla}, G., {Mandel}, K., \& {Sohn}, S.~T. 2018, \apj, 857, 78,
  \dodoi{10.3847/1538-4357/aab78f}

\bibitem[{{Patel} {et~al.}(2017{\natexlab{b}}){Patel}, {Besla}, \&
  {Sohn}}]{Patel2017a}
{Patel}, E., {Besla}, G., \& {Sohn}, S.~T. 2017{\natexlab{b}}, \mnras, 464,
  3825, \dodoi{10.1093/mnras/stw2616}

\bibitem[{{Pe{\~n}arrubia} \& {Fattahi}(2017)}]{Penarrubia2017}
{Pe{\~n}arrubia}, J., \& {Fattahi}, A. 2017, \mnras, 468, 1300,
  \dodoi{10.1093/mnras/stx323}

\bibitem[{{Pe{\~n}arrubia} {et~al.}(2016){Pe{\~n}arrubia}, {G{\'o}mez},
  {Besla}, {Erkal}, \& {Ma}}]{Penarrubia2016}
{Pe{\~n}arrubia}, J., {G{\'o}mez}, F.~A., {Besla}, G., {Erkal}, D., \& {Ma},
  Y.-Z. 2016, \mnras, 456, L54, \dodoi{10.1093/mnrasl/slv160}

\bibitem[{{Pe{\~n}arrubia} {et~al.}(2014){Pe{\~n}arrubia}, {Ma}, {Walker}, \&
  {McConnachie}}]{Penarrubia2014}
{Pe{\~n}arrubia}, J., {Ma}, Y.-Z., {Walker}, M.~G., \& {McConnachie}, A. 2014,
  \mnras, 443, 2204, \dodoi{10.1093/mnras/stu879}

\bibitem[{{Peebles}(2017)}]{Peebles:2017}
{Peebles}, P.~J.~E. 2017, arXiv e-prints, arXiv:1705.10683.
\newblock \doarXiv{1705.10683}

\bibitem[{P\'erez \& Granger(2007)}]{ipython}
P\'erez, F., \& Granger, B.~E. 2007, Computing in Science and Engineering, 9,
  21, \dodoi{10.1109/MCSE.2007.53}

\bibitem[{{Petersen} \& {Pe{\~n}arrubia}(2020)}]{Petersen:2020}
{Petersen}, M.~S., \& {Pe{\~n}arrubia}, J. 2020, \mnras, 494, L11,
  \dodoi{10.1093/mnrasl/slaa029}

\bibitem[{{Petersen} \& {Pe{\~n}arrubia}(2021)}]{Petersen2021}
---. 2021, Nature Astronomy, 5, 251, \dodoi{10.1038/s41550-020-01254-3}

\bibitem[{{Planck Collaboration} {et~al.}(2020){Planck Collaboration},
  {Aghanim}, {Akrami}, {Ashdown}, {Aumont}, {Baccigalupi}, {Ballardini},
  {Banday}, {Barreiro}, {Bartolo}, {Basak}, {Battye}, {Benabed}, {Bernard},
  {Bersanelli}, {Bielewicz}, {Bock}, {Bond}, {Borrill}, {Bouchet}, {Boulanger},
  {Bucher}, {Burigana}, {Butler}, {Calabrese}, {Cardoso}, {Carron},
  {Challinor}, {Chiang}, {Chluba}, {Colombo}, {Combet}, {Contreras}, {Crill},
  {Cuttaia}, {de Bernardis}, {de Zotti}, {Delabrouille}, {Delouis}, {Di
  Valentino}, {Diego}, {Dor{\'e}}, {Douspis}, {Ducout}, {Dupac}, {Dusini},
  {Efstathiou}, {Elsner}, {En{\ss}lin}, {Eriksen}, {Fantaye}, {Farhang},
  {Fergusson}, {Fernandez-Cobos}, {Finelli}, {Forastieri}, {Frailis},
  {Fraisse}, {Franceschi}, {Frolov}, {Galeotta}, {Galli}, {Ganga},
  {G{\'e}nova-Santos}, {Gerbino}, {Ghosh}, {Gonz{\'a}lez-Nuevo}, {G{\'o}rski},
  {Gratton}, {Gruppuso}, {Gudmundsson}, {Hamann}, {Handley}, {Hansen},
  {Herranz}, {Hildebrandt}, {Hivon}, {Huang}, {Jaffe}, {Jones}, {Karakci},
  {Keih{\"a}nen}, {Keskitalo}, {Kiiveri}, {Kim}, {Kisner}, {Knox},
  {Krachmalnicoff}, {Kunz}, {Kurki-Suonio}, {Lagache}, {Lamarre}, {Lasenby},
  {Lattanzi}, {Lawrence}, {Le Jeune}, {Lemos}, {Lesgourgues}, {Levrier},
  {Lewis}, {Liguori}, {Lilje}, {Lilley}, {Lindholm}, {L{\'o}pez-Caniego},
  {Lubin}, {Ma}, {Mac{\'\i}as-P{\'e}rez}, {Maggio}, {Maino}, {Mandolesi},
  {Mangilli}, {Marcos-Caballero}, {Maris}, {Martin}, {Martinelli},
  {Mart{\'\i}nez-Gonz{\'a}lez}, {Matarrese}, {Mauri}, {McEwen}, {Meinhold},
  {Melchiorri}, {Mennella}, {Migliaccio}, {Millea}, {Mitra},
  {Miville-Desch{\^e}nes}, {Molinari}, {Montier}, {Morgante}, {Moss}, {Natoli},
  {N{\o}rgaard-Nielsen}, {Pagano}, {Paoletti}, {Partridge}, {Patanchon},
  {Peiris}, {Perrotta}, {Pettorino}, {Piacentini}, {Polastri}, {Polenta},
  {Puget}, {Rachen}, {Reinecke}, {Remazeilles}, {Renzi}, {Rocha}, {Rosset},
  {Roudier}, {Rubi{\~n}o-Mart{\'\i}n}, {Ruiz-Granados}, {Salvati}, {Sandri},
  {Savelainen}, {Scott}, {Shellard}, {Sirignano}, {Sirri}, {Spencer},
  {Sunyaev}, {Suur-Uski}, {Tauber}, {Tavagnacco}, {Tenti}, {Toffolatti},
  {Tomasi}, {Trombetti}, {Valenziano}, {Valiviita}, {Van Tent}, {Vibert},
  {Vielva}, {Villa}, {Vittorio}, {Wandelt}, {Wehus}, {White}, {White},
  {Zacchei}, \& {Zonca}}]{Planck2018}
{Planck Collaboration}, {Aghanim}, N., {Akrami}, Y., {et~al.} 2020, \aap, 641,
  A6, \dodoi{10.1051/0004-6361/201833910}

\bibitem[{Price-Whelan(2017)}]{gala}
Price-Whelan, A.~M. 2017, The Journal of Open Source Software, 2,
  \dodoi{10.21105/joss.00388}

\bibitem[{{Putman} {et~al.}(2021){Putman}, {Zheng}, {Price-Whelan}, {Grcevich},
  {Johnson}, {Tollerud}, \& {Peek}}]{Putman:2021}
{Putman}, M.~E., {Zheng}, Y., {Price-Whelan}, A.~M., {et~al.} 2021, \apj, 913,
  53, \dodoi{10.3847/1538-4357/abe391}

\bibitem[{{Salomon} {et~al.}(2021){Salomon}, {Ibata}, {Reyl{\'e}}, {Famaey},
  {Libeskind}, {McConnachie}, \& {Hoffman}}]{Salomon2021}
{Salomon}, J.~B., {Ibata}, R., {Reyl{\'e}}, C., {et~al.} 2021, \mnras, 507,
  2592, \dodoi{10.1093/mnras/stab2253}

\bibitem[{{Salvatier} {et~al.}(2016){Salvatier}, {Wiecki$\hat{\rm a}$}, \&
  {Fonnesbeck}}]{Salvatier2016}
{Salvatier}, J., {Wiecki$\hat{\rm a}$}, T.~V., \& {Fonnesbeck}, C. 2016,
  {PyMC3: Python probabilistic programming framework}.
\newblock \doeprint{1610.016}

\bibitem[{{Sawala} {et~al.}(2022){Sawala}, {McAlpine}, {Jasche}, {Lavaux},
  {Jenkins}, {Johansson}, \& {Frenk}}]{Sawala2022}
{Sawala}, T., {McAlpine}, S., {Jasche}, J., {et~al.} 2022, \mnras, 509, 1432,
  \dodoi{10.1093/mnras/stab2684}

\bibitem[{{Sch{\"o}nrich} {et~al.}(2010){Sch{\"o}nrich}, {Binney}, \&
  {Dehnen}}]{Schonrich2010}
{Sch{\"o}nrich}, R., {Binney}, J., \& {Dehnen}, W. 2010, \mnras, 403, 1829,
  \dodoi{10.1111/j.1365-2966.2010.16253.x}

\bibitem[{{Sohn} {et~al.}(2012){Sohn}, {Anderson}, \& {van der
  Marel}}]{Sohn:2012}
{Sohn}, S.~T., {Anderson}, J., \& {van der Marel}, R.~P. 2012, \apj, 753, 7,
  \dodoi{10.1088/0004-637X/753/1/7}

\bibitem[{{Tolstoy} {et~al.}(2009){Tolstoy}, {Hill}, \& {Tosi}}]{Tolstoy:2009}
{Tolstoy}, E., {Hill}, V., \& {Tosi}, M. 2009, \araa, 47, 371,
  \dodoi{10.1146/annurev-astro-082708-101650}

\bibitem[{{van der Marel} {et~al.}(2012){van der Marel}, {Fardal}, {Besla},
  {Beaton}, {Sohn}, {Anderson}, {Brown}, \& {Guhathakurta}}]{vdm2012}
{van der Marel}, R.~P., {Fardal}, M., {Besla}, G., {et~al.} 2012, \apj, 753, 8,
  \dodoi{10.1088/0004-637X/753/1/8}

\bibitem[{{van der Marel} {et~al.}(2019){van der Marel}, {Fardal}, {Sohn},
  {Patel}, {Besla}, {del Pino}, {Sahlmann}, \& {Watkins}}]{vdm2019}
{van der Marel}, R.~P., {Fardal}, M.~A., {Sohn}, S.~T., {et~al.} 2019, \apj,
  872, 24, \dodoi{10.3847/1538-4357/ab001b}

\bibitem[{{van der Marel} \& {Guhathakurta}(2008)}]{vdm2008}
{van der Marel}, R.~P., \& {Guhathakurta}, P. 2008, \apj, 678, 187,
  \dodoi{10.1086/533430}

\bibitem[{{Villanueva-Domingo} {et~al.}(2021){Villanueva-Domingo},
  {Villaescusa-Navarro}, {Genel}, {Angl{\'e}s-Alc{\'a}zar}, {Hernquist},
  {Marinacci}, {Spergel}, {Vogelsberger}, \&
  {Narayanan}}]{Villanueva-Domingo2021}
{Villanueva-Domingo}, P., {Villaescusa-Navarro}, F., {Genel}, S., {et~al.}
  2021, arXiv e-prints, arXiv:2111.14874.
\newblock \doarXiv{2111.14874}

\bibitem[{Virtanen {et~al.}(2020)Virtanen, Gommers, Oliphant, Haberland, Reddy,
  Cournapeau, Burovski, Peterson, Weckesser, Bright, {van der Walt}, Brett,
  Wilson, Millman, Mayorov, Nelson, Jones, Kern, Larson, Carey, Polat, Feng,
  Moore, {VanderPlas}, Laxalde, Perktold, Cimrman, Henriksen, Quintero, Harris,
  Archibald, Ribeiro, Pedregosa, {van Mulbregt}, \& {SciPy 1.0
  Contributors}}]{scipy}
Virtanen, P., Gommers, R., Oliphant, T.~E., {et~al.} 2020, Nature Methods, 17,
  261, \dodoi{10.1038/s41592-019-0686-2}

\bibitem[{{Wang} {et~al.}(2022){Wang}, {Hammer}, \& {Yang}}]{Wang:2022}
{Wang}, J., {Hammer}, F., \& {Yang}, Y. 2022, \mnras, 510, 2242,
  \dodoi{10.1093/mnras/stab3258}

\bibitem[{Watkins {et~al.}(2010)Watkins, Evans, \& An}]{Watkins2010}
Watkins, L.~L., Evans, N.~W., \& An, J.~H. 2010, Monthly Notices of the Royal
  Astronomical Society, 406, 264, \dodoi{10.1111/j.1365-2966.2010.16708.x}

\bibitem[{{Zhai} {et~al.}(2020){Zhai}, {Guo}, {Zhao}, {Gu}, \&
  {Liu}}]{Zhai2020}
{Zhai}, M., {Guo}, Q., {Zhao}, G., {Gu}, Q., \& {Liu}, A. 2020, \apj, 890, 27,
  \dodoi{10.3847/1538-4357/ab6986}

\end{thebibliography}
\bibliographystyle{aasjournal}

\end{document}